\documentclass[11pt]{article}
\usepackage{times}
\usepackage{graphicx}
\usepackage{multirow}
\usepackage{amssymb,amsmath}
\usepackage{color}
\usepackage{fullpage}
\usepackage{url}
\usepackage[ruled, vlined]{algorithm2e} 

\usepackage[square,sort&compress,comma,numbers]{natbib}
\newcommand{\topic}[1]{\smallskip  \noindent{\underline{\bf #1:}}}

\usepackage{epsfig, subfigure}
\long\def\ignore#1{}
\newtheorem{theorem}{Theorem}
\newtheorem{definition}{Definition}

\newtheorem{example}{Example}
\newenvironment{proof}{\noindent\textbf{Proof: }\ignorespaces}{}
\newcommand{\qed}{\hspace*{\fill}$\Box$\medskip}

\newcommand{\calT}{\mathcal{T}}
\newcommand{\calY}{\mathcal{Y}}

\newcommand{\calF}{\mathcal{F}}

\newcommand{\Prob}{\Pr}
\newcommand{\calX}{\mathcal{X}}

\newcommand{\rank}{\Upsilon}

\newcommand{\bfe}{\mathbf{e}}
\newcommand{\bfc}{\mathbf{c}}
\newcommand{\bfu}{\mathbf{u}}
\newcommand{\bfd}{\mathbf{d}}
\newcommand{\bfi}{\mathbf{i}}
\newcommand{\bfF}{\mathbf{F}}

\restylealgo{boxruled}
\dontprintsemicolon

\newcommand{\rmk}{k}
\newcommand{\Exp}{\operatorname{E}}
\newcommand{\eat}[1]{}
\newcommand{\eatforjdmr}[1]{}
\newcommand{\uimag}{\jmath}
\newcommand{\barw}{\bar\omega}
\newcommand{\tilw}{\tilde\omega}

\newcounter{foo}

\newenvironment{mylist}{\begin{list}{$\bullet$}
    {   \setlength{\itemsep}{0pt}
        \setlength{\topsep}{0pt}}
    }
{\end{list}}

\newcommand{\omitforsigmod}[1]{}  

\newcommand{\Cvee}{\textcircled{\small{$\vee$}}}
\newcommand{\Cwedge}{\textcircled{\small{$\wedge$}}}
\newcommand{\Topk}{\emph{Top-$k$}}
\renewcommand{\Topk}{top-k}
\newcommand{\UTK}{$\mathsf{U}$-$\mathsf{Top}$}
\newcommand{\URK}{$\mathsf{U}$-$\mathsf{Rank}$}
\newcommand{\PRF}{$\mathsf{PRF}$}
\newcommand{\PRFs}{$\mathsf{PRF^\omega}$}
\newcommand{\PRFsPARAM}[1]{$\mathsf{PRF^\omega}(#1)$}
\newcommand{\PRFe}{$\mathsf{PRF^e}$}
\newcommand{\ERK}{$\mathsf{E}$-$\mathsf{Rank}$}
\newcommand{\PT}{$\mathsf{PT}(h)$}
\newcommand{\PTPARAM}[1]{$\mathsf{PT}(#1)$}
\newcommand{\ES}{$\mathsf{E}$-$\mathsf{Score}$}
\newcommand{\CON}{$\mathsf{Con}$-$\mathsf{Topk}$}
\newcommand{\dist}{\mathsf{dis}}
\newcommand{\score}{\mathsf{score}}
\newcommand{\rk}{r}
\newcommand{\PRFl}{$\mathsf{PRF^\ell}$}

\newcommand{\PS}{P}

\title{A Unified Approach to Ranking in Probabilistic Databases}

\author{Jian Li
\thanks{Computer Science Department, University of Maryland, College Park 20742, MD, USA. Email: lijian@cs.umd.edu}
 \and Barna Saha
\thanks{Computer Science Department, University of Maryland, College Park 20742, MD, USA. Email: barna@cs.umd.edu}
  \and Amol Deshpande 
\thanks{Computer Science Department, University of Maryland, College Park 20742, MD, USA. Email: amol@cs.umd.edu}
}

\date{}

\begin{document}

\maketitle
\begin{abstract}
Ranking is a fundamental operation in data analysis and decision support, and plays an even
more crucial role if the dataset being explored exhibits uncertainty. This
has led to much work in understanding how to rank the tuples in a probabilistic dataset in recent years.
In this article, we present a unified approach to ranking and top-$k$ query processing in
probabilistic databases by viewing it as a multi-criteria optimization problem, and by deriving a
set of {\em features} that capture the key properties of a probabilistic dataset that dictate
the ranked result. We contend that a single, specific ranking function may not suffice for probabilistic
databases, and we instead propose two {\em parameterized ranking functions}, called \PRFs\ and \PRFe, that generalize or
can approximate many of the previously proposed ranking functions. We present novel
{\em generating functions}-based algorithms for
efficiently ranking large datasets according to these ranking functions, even if the
datasets exhibit complex correlations modeled using {\em probabilistic and/xor trees}
or {\em Markov networks}. We further propose that the parameters of the ranking function be
{\em learned} from user preferences, and we develop an approach to learn those parameters.
Finally, we present a comprehensive experimental study that illustrates the effectiveness of
our parameterized ranking functions, especially \PRFe, at approximating other ranking functions and
the scalability of our proposed algorithms for exact or approximate ranking.
\end{abstract}

\section{Introduction}
\label{sec_introduction}
Recent years have seen a dramatic increase in the number of applications domains that naturally generate
uncertain data and that demand support for executing complex  decision support queries over them.
These include
information retrieval~\cite{fuhr:is97}, data integration and cleaning~\cite{andritsos:icde06,conf/vldb/DongHY07}, text
analytics~\cite{journals/debu/JayramKRVZ06,gupta06}, social network
analysis~\cite{journals/debu/AdarR07}, sensor data management~\cite{DGM2005,cheng:sigmod03}, financial applications, biological and scientific data management, etc.
Uncertainty arises in these environments for a variety of reasons.
Sensor data typically contains noise and measurement errors, and is often
incomplete because of sensor faults or communication link failures. In social networks and
scientific domains, the observed interaction or experimental data is often very noisy, and ubiquitous
use of predictive models adds
a further layer of uncertainty. Use of automated tools in data
integration and information extraction can introduce significant uncertainty in the output.

By their very nature, many of these applications require support for 
{\em ranking} or {\em top-k query processing} over large volumes of data. For instance, consider a {\em House Search} application where a user
is searching for a house using a real estate sales dataset that lists the houses for sale. 
Such a dataset, which may be constructed by crawling and combining data from multiple
sources, is inherently uncertain and noisy. In fact, the houses that the user prefers the most, are
also the most likely to be sold by now. We may denote such uncertainty by associating with
each advertisement a {\em probability} that it is still valid. Incorporating such uncertainties
into the returned answers is, however, a challenge considering the complex interplay between the relevance of a
house by itself, and the probability that the advertisement is still valid.

Many other application domains also exhibit resource constraints of some form, and we must somehow rank the entities
or tuples under consideration to select the most relevant objects to focus our attention on. For example,
in financial applications, we may want to choose the best
stocks in which to invest, given their expected performance in the future (which is uncertain at best).
In learning or classification tasks, we often need to choose the best ``k'' features to use~\cite{conf/uai/ZukED07}.
In sensor networks or scientific databases, we may not know the ``true'' values of the physical properties being measured because
of measurement noises or failures~\cite{DGM2005}, but we may still need to choose a set of sensors or entities
in response to a user query.

Ranking in presence of uncertainty is non-trivial even if the relevance scores can be computed easily (the
main challenge in the deterministic case),
mainly because of the complex trade-offs introduced by the score distributions and the tuple uncertainties.
This has led to many ranking functions being proposed for combining the scores and the probabilities in recent years,
all of which appear quite natural at the first glance (we review several of them in detail later).
We begin with a systematic
exploration of these issues by recognizing that ranking in probabilistic databases is inherently a multi-criteria
optimization problem, and by deriving a set of {\em features}, the key properties of a probabilistic
dataset that influence the ranked result. We empirically illustrate the diverse and conflicting
behavior of several natural ranking functions, and argue that a single specific ranking function may
not be appropriate to rank different uncertain databases that we may encounter in practice. Furthermore, different
users may weigh the features differently, resulting in different rankings over the same dataset.
We then define a general and powerful ranking function,
called {\em \PRF}, that allows us to explore the space of possible ranking functions. We discuss
its relationship to previously proposed ranking functions, and also identify two specific parameterized
ranking functions, called \PRFs\ and \PRFe, as being interesting. The \PRFs\ ranking function is
essentially a linear, weighted ranking function that resembles the scoring functions typically used in
information retrieval, web search, data integration, keyword query answering
etc.~\cite{Herbrich98,Joachims02,RankNet,Dekel04,Talukdar08}.
We observe that \PRFs\ may not be suitable for ranking large datasets due to its high running time,
and instead propose \PRFe, which uses a single parameter, and can effectively approximate previously proposed ranking functions
for probabilistic databases very well.

We then develop novel algorithms based on {\em generating functions} to efficiently rank the tuples
in a probabilistic dataset using any \PRF\ ranking function. Our algorithm can handle a
probabilistic dataset with arbitrary correlations; however, it is particularly efficient when the
probabilistic database contains only {\em mutual exclusivity} and/or {\em co-existence}
correlations (called {\em probabilistic and/xor trees}~\cite{pods09_LD}).
Our main contributions can be summarized as follows:
\begin{list}{$\bullet$}{\leftmargin 0.10in \topsep 2pt \itemsep 2pt}
        \item We develop a framework for {\em learning} ranking functions over probabilistic databases by
            identifying a set of key {\em features}, by proposing several parameterized ranking functions over
            those features, and by choosing the parameters based on user preferences or feedback.
    \item We present novel algorithms based on {\em generating functions} that enable
        us to efficiently rank very large datasets. Our key algorithm is an
        $O(n \log(n))$ algorithm for evaluating a \PRFe\
        function over datasets with low correlations (specifically, constant height probabilistic and/xor trees). The algorithm runs in $O(n)$ time if the dataset is pre-sorted by score.
   \item We present a polynomial time algorithm for ranking a correlated dataset when the correlations are captured using a bounded-treewidth graphical model. The algorithm we present is actually
    for computing the probability that a given tuple is ranked at a given position across all the possible worlds, and is of independent interest.
    \item We develop a novel, DFT-based algorithm for approximating an arbitrary weighted ranking function using a linear combination
    of \PRFe\ functions.
\item We show that a \PRFs\ ranked result can be seen as a {\em consensus} answer under a suitably defined distance function -- a consensus answer
    is defined to be the answer that is closest in expectation to the answers over the possible worlds.
   %
    \item We present a comprehensive experimental study over several real and synthetic datasets, comparing the behavior of the ranking functions and
    the effectiveness of our proposed algorithms.
\end{list}

\topic{Outline} We begin with a brief discussion of the related work (Section \ref{sec:related
work}). In Section \ref{sec:problem formulation}, we review our probabilistic database model and
the prior work on ranking in probabilistic databases, and propose two parameterized ranking functions. In Section \ref{sec:algorithms},
we present our generating functions-based algorithms for ranking. We then present an approach to approximate
different ranking functions using our parameterized ranking functions, and to learn a ranking function from user preferences (Section \ref{sec:approximating and learning}).
In Section \ref{sec:contop-k}, we explore the connection between \PRFs\ and consensus top-k query results.
In Section \ref{sec:prfeprop}, we observe an interesting property of the \PRFe\ function that helps us gain better insight into
its behavior. We then present a comprehensive experiment study in Section \ref{sec:experiments}.
Finally, in Section \ref{sec:correlations}, we develop an algorithm for handling correlated datasets where the correlations are
captured using bounded-treewidth graphical models.

\section{Related Work}
\label{sec:related work}
There has been much work on managing probabilistic, uncertain, incomplete, and/or fuzzy data in
database systems (see, e.g., \cite{lakshmanan:tods97,fuhr:is97,cheng:sigmod03,dalvi:vldb04,widom:cidr05,Koch09,GT06_models}).
The work in this area has spanned
a range of issues from theoretical development of data models and data languages to practical
implementation issues such as indexing techniques; several research efforts are underway to build
systems to manage uncertain data (e.g., MYSTIQ~\cite{dalvi:vldb04}, Trio~\cite{widom:cidr05},
ORION~\cite{cheng:sigmod03}, MayBMS~\cite{Koch09}, PrDB~\cite{sen:vldbj09}).
The approaches can be differentiated based on whether they support
{\em tuple-level uncertainty} where ``existence'' probabilities are attached to the tuples of the
database, or {\em attribute-level uncertainty} where (possibly continuous) probability
distributions are attached to the attributes, or both. The proposed approaches differ further based on whether
they consider correlations or not. Most work in probabilistic databases has either assumed
independence~\cite{fuhr:is97,dalvi:vldb04} or has restricted the correlations that can be
modeled~\cite{lakshmanan:tods97,andritsos:icde06,sarma:icde06}. More recently, several
approaches have been presented that allow representation of arbitrary correlations and querying over correlated databases~\cite{GT06_models,sen:vldbj09,Koch08}.

The area of ranking and top-k query processing has also seen much work in databases (see, e.g., Ilyas et
al.'s survey~\cite{journals/csur/IlyasBS08}). More recently, several researchers have considered
top-k query processing in probabilistic databases. Soliman et al.~\cite{soliman:icde07} defined the
problem of ranking over probabilistic databases, and proposed two ranking functions to combine
tuple scores and probabilities.  Yi et al.~\cite{conf/icde/YiLSK08} present improved algorithms for
the same ranking functions.
Zhang and Chomicki~\cite{conf/dbrank/ZhangC08}
present a desiderata for ranking functions, and propose the notion of {\em Global Top-k} answers. Ming Hua et al.~\cite{conf/sigmod/HuaPZL08}
propose {\em probabilistic threshold ranking}, which is quite similar to Global Top-k.
Cormode et al.~\cite{Cormode09} also present a semantics of ranking functions and
a new ranking function called {\em expected rank}.
Liu et al.~\cite{conf/dasfaa/LiuYXTL10} propose the notion of {\em k}-selection queries; unlike
most of the above definitions, the result here is sensitive to the actual tuple scores.
We will review these ranking functions in detail in next section.
Ge et al.~\cite{ge2009top} propose the notion of
{\em typical answers}, where they propose returning a collection of typical answers instead
of just one answer. This can be seen as complementary to our approach here; one could show the
typical answers to the user to understand the user preferences during an exploratory phase, and
then learn a single ranking function to rank using the techniques developed in this article.

There has also been work on top-k query processing in probabilistic databases where
the ranking is by the result tuple {\em probabilities} (i.e., probability and score are
identical)~\cite{conf/icde/ReDS07}.
The main challenge in that work is efficient computation of the
probabilities, whereas we assume that the probability and score are either given or can be computed
easily.

The aforementioned work has focused mainly on tuple uncertainty and discrete attribute uncertainty.
Soliman and Ilyas~\cite{soliman2009ranking} were the first to consider the problem of handling continuous distributions.
Recently, in a followup work~\cite{conf/vldb/Li10ranking},
we extended the algorithm for \PRF\ to arbitrary continuous distributions.
We were able to obtain exact polynomial time algorithms for some continuous probability distribution classes,
and efficient approximation schemes with provable guarantees for arbitrary probability distributions.
One important ingredient of those algorithms is an extension of the generating function used in this article.

Recently, there has also been much work on nearest neighbor-style queries over
uncertain datasets~\cite{dasfaa07-nn,icde08-probnn,vldb08-knn,edbt09-knn}.
In fact, a nearest neighbor query (or a $k$-nearest neighbor query) can be seen as a ranking
query where the score of a point is the distance of that point to the given query point.
Thus, our new ranking semantics and algorithms can be directly used for nearest neighbor queries
over uncertain points with discrete probability distributions.

There is a tremendous body of work on ranking documents in information retrieval, and learning how to rank
documents given user preferences (see Liu~\cite{liu2009learning} for a comprehensive survey). That work has considered
aspects such as different ranking models, loss functions, different scoring techniques etc. The techniques developed there
tend to be specific to document retrieval (focusing on keywords, terms, and relevance), and usually do not deal with
{\em existence} uncertainty (although they often do model document relevance as a random variable). Furthermore, our work here primarily
focuses on highly efficient algorithms for ranking using a spectrum of different ranking functions. Exploring and understanding the
connections between the two research areas is a fruitful direction for further research.


Finally, we note that one \PRF\ function is only able to model preferences of one user.
There is an increasing interest in finding a ranking that satisfies multiple users having diverse preferences and intents.
Several new theoretical models have been proposed recently~\cite{azar2009multiple,bansal2010approximation, azar2010ranking}.
However, all the inputs are assumed to be certain in those models. Incorporating uncertainty into those models or introducing
the notion of diversity into our model is an interesting research direction.

\begin{figure*}[t]
\begin{minipage}[t]{0.37\linewidth}%
    \scriptsize{
        \begin{tabular}{|c|c|c|c|c|c|c|c|}
        \hline
        Time & Car &  Plate & Speed & \dots & Prob & Tuple \\
             & Loc &   No   &  & &  &  Id \\
        \hline
        11:40 &L1& X-123 & 120 &  \dots & 0.4 & $t_1$\\
        \hline
        11:55 &L2& Y-245 & 130 &  \dots &0.7 & $t_2$\\
        \hline
        11:35 &L3& Y-245 & 80 & \dots & 0.3 & $t_3$ \\
        \hline
        12:10 &L4&  Z-541 & 95 &  \dots &0.4 & $t_4$\\
        \hline
        12:25 &L5&  Z-541 & 110 & \dots & 0.6 & $t_5$\\
        \hline
        12:15 &L6& L-110& 105 & \dots & 1.0 & $t_6$\\
        \hline
        \end{tabular}
    }
\end{minipage}
\begin{minipage}[t]{0.19\linewidth}%
\scriptsize{
\begin{tabular}{|c|c|}
\hline
Possible Worlds & Prob \\ \hline
$pw_{1}=\{t_2,t_1,t_6,t_4\}$ & .112 \\
$pw_{2}=\{t_2,t_1,t_5,t_6\}$ & .168 \\
$pw_{3}=\{t_1,t_6,t_4,t_3\}$ & .048\\
$pw_{4}=\{t_1,t_5,t_6,t_3\}$ & .072\\
$pw_{5}=\{t_2,t_6,t_4\}$ & .168\\
$pw_{6}=\{t_2,t_5,t_6\}$ & .252\\
$pw_{7}=\{t_6,t_4,t_3\}$ & .072\\
$pw_{8}=\{t_5,t_6,t_3\}$ & .108\\
\hline
\end{tabular}
}
\end{minipage}
\mbox{\ \ \ \ \ \ \ \ }
\begin{minipage}{0.31\linewidth}%
{\includegraphics[width=2.5in]{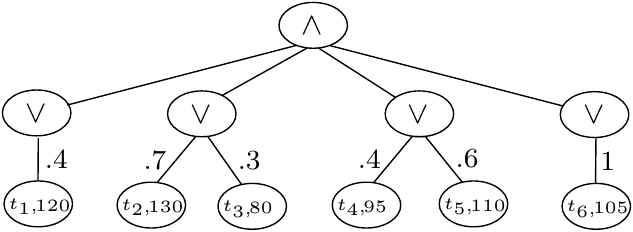}}
\end{minipage}
\caption{Example of a probabilistic database which contains automatically captured information about
speeding cars -- here the {\em Plate No.} is the possible worlds key and the {\em speed} is the score attribute
that we will use for ranking.
Tuples $t_2$ and $t_3$ (similarly, $t_4$ and $t_5$) are mutually exclusive.
The second table lists all possible worlds. Note that the tuples are sorted according
to their speeds in each possible world.
The corresponding and/xor tree compactly encodes these correlations.}
\label{fig_example}
\end{figure*}

\section{Problem Formulation}
\label{sec:ranking}
\label{sec:problem formulation}
We begin with defining our model of a probabilistic database,
called {\em probabilistic and/xor tree}~\cite{pods09_LD}, that
captures several common types of correlations.
We then review the prior work on top-k query processing
in probabilistic databases, and argue that a single specific ranking function
may not capture the intricacies of ranking with
uncertainty. We then present our parameterized ranking functions, \PRFs\ and \PRFe.

\subsection{Probabilistic Database Model}
\label{sec:preliminaries}
We use the prevalent {\em possible worlds semantics} for probabilistic
databases~\cite{dalvi:vldb04}. We denote a probabilistic relation with tuple uncertainty by $D_T$,
where $T$ denotes the set of tuples (in Section \ref{sec:attribute_uncertainty}, we present extensions to handle attribute uncertainty).  The set of all possible worlds is denoted by $PW=\{pw_1,
pw_2,...., pw_n\}$. Each tuple $t_i \in T$ is associated with
an existence probability $\Prob(t_i)$ and a score $\score(t_i)$,
computed based on a scoring function $\score: T\rightarrow \mathbb{R}$.  Usually $\score(t)$
is computed based on the tuple attribute values and measures the relative user preference for different
tuples. In a deterministic database, tuples with higher scores should be ranked higher.
We use $r_{pw}:T\rightarrow \{1,\ldots, n\}\cup \{\infty\}$ to denote the rank of the tuple
$t$ in a possible world $pw$ according to $\score$.  If $t$ does not appear in the possible world $pw$,
we let $r_{pw}(t)=\infty$.  We say $t_1$ {\em ranks higher} than $t_2$ in the possible world $pw$ if
$r_{pw}(t_1) < r_{pw}(t_2)$.
For each tuple $t$, we define a random variable $r(t)$ that denotes the rank of $t$ in $D_T$.

\begin{definition}
The {\em positional probability} of a tuple $t$ being ranked at position $k$, denoted $\Prob(r(t) = k)$, is
the total probability of the possible worlds where $t$ is ranked at position $k$.
The {\em rank distribution} of a tuple $t$, denoted $\Prob(r(t))$, is simply the probability distribution of the random variable $r(t)$.
\end{definition}

\topic{Probabilistic And/Xor Tree Model}
Our algorithms can handle arbitrarily correlated relations where correlations
modeled using Markov networks (Section \ref{sec:correlations}). However, in most of
this article, we focus on the {\em probabilistic and/xor tree model}, introduced in our prior work~\cite{pods09_LD}, that
can capture only a more restricted set of correlations, but admits
highly efficient query processing algorithms. More specifically, an and/xor tree
captures two types of correlations: (1) {\em mutual exclusivity} (denoted $\Cvee$ ({\em xor}))
and (2) {\em mutual co-existence} ($\Cwedge$ ({\em and})). Two events satisfy the mutual co-existence
correlation if, in any possible world, either both events occur or neither
occurs. Similarly two events are mutually exclusive if there is no possible world where both
happen. 

Now, let us formally define a probabilistic and/xor tree. In tree $\calT$, we denote the set of children of node $v$ by $Ch_{\calT}(v)$ and
the least common ancestor of two leaves $l_1$ and $l_2$
by $LCA_{\calT}(l_1,l_2)$. We omit the subscript if the context is clear.
For simplicity, we separate the attributes of the relation into two groups: (1) a possible worlds {\em key}, denoted $K$, which is
unique in any possible world (i.e., two tuples that agree on $K$ are mutually exclusive), and (2) the value attributes, denoted $A$.
If the relation does not have any key attributes, $K = \phi$.

\begin{definition}
\label{and/xor}
A probabilistic and/xor tree $\calT$ represents the mutual exclusion and co-existence
correlations in a probabilistic relation $R^P(K; A)$,
where $K$ is the possible worlds key, and $A$ denotes the value attributes.
In $\calT$, each leaf denotes a tuple,
and each inner node has a mark, $\Cvee$ or $\Cwedge$.
For each $\Cvee$ node $u$ and each of its children $v\in Ch(u)$,
there is a nonnegative value $p_{(u,v)}$ associated with the edge $(u,v)$.
Moreover, we require:
\begin{list}{$\bullet$}{\leftmargin 0.25in \topsep 3pt \itemsep 2pt}
\item (Probability Constraint) $\sum_{v:v\in Ch(u)}\Prob(u,v)\leq 1$.
\item (Key Constraint) For any two different leaves $l_1,l_2$ holding the same key,
$LCA(l_1,l_2)$ is a $\Cvee$ node\footnote{
The key constraint is imposed to avoid two leaves with the same key
but different attribute values coexisting in a possible world.
}.
\end{list}
Let $\calT_v$ be the subtree rooted at $v$ and $Ch(v)=\{v_1,\ldots,v_\ell\}$.
The subtree $\calT_v$ inductively defines a random subset $S_v$ of its leaves
by the following independent process:
\begin{list}{$\bullet$}{\leftmargin 0.25in \topsep 2pt \itemsep 1pt}
\item
If $v$ is a leaf, $S_v=\{v\}$.
\item
If $\calT_v$ roots at a $\Cvee$ node,
then \\ \mbox{\ \ \ \ \ \ \ } $S_v= \left\{
             \begin{array}{ll}
               S_{v_i}         & \hbox{with prob $p_{(v,v_i)}$} \\
               \emptyset       & \hbox{with prob $1-\sum_{i=1}^\ell p_{(v,v_i)}$}
             \end{array}
           \right.
$
\item
If $\calT_v$ roots at a $\Cwedge$ node,
then $S_v=  \cup_{i=1}^\ell S_{v_i}$
\end{list}
\end{definition}


\begin{figure}[t]
\begin{minipage}[t]{\linewidth}%
\center{
\begin{tabular}{|l|c|}
\hline
Possible Worlds & Prob \\ \hline
$pw_{1}=\{(t_3,6), (t_2,5), (t_11)\}$ & .3 \\
$pw_{2}=\{(t_3,9), (t_1,7)\}$ & .3 \\
$pw_{3}=\{(t_2,8),(t_4,4),(t_5,3)\}$ & .4\\
\hline
\end{tabular}
}
\vspace{0.2cm}
\end{minipage}
\begin{minipage}{\linewidth}%
\center{\includegraphics[width=0.7\linewidth]{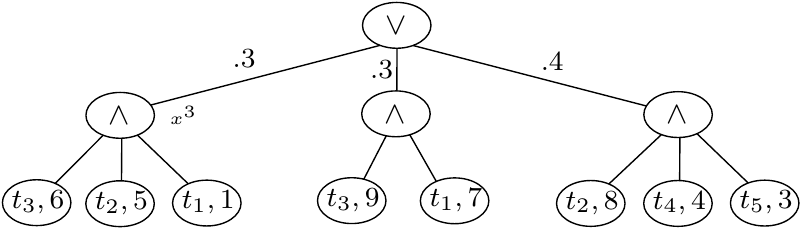}}
\end{minipage}
\caption{
Example of a highly correlated probabilistic database with $3$ possible worlds
and the and/xor tree that captures the correlation.
}
\label{eg_possibleworld}
\vspace{-0.2cm}
\end{figure}

{\em x-tuples} (which can be used to specify mutual exclusivity correlations between tuples)
correspond to the special case where we have a tree of {\em height} 2, with a $\Cwedge$ node as the root and
only $\Cvee$ nodes in the second level.
Figure \ref{eg_possibleworld} shows an example of
an and/xor tree that models the data from a traffic monitoring
application~\cite{soliman:icde07}, where the tuples represent
automatically captured traffic data.
The inherent uncertainty in the
monitoring infrastructure is captured using an and/xor tree, that
encodes the tuple existence probabilities as well as the
correlations between the tuples.
For example, the leftmost $\Cvee$ node
indicates $t_1$ is present with probability $.4$ and the
second $\Cvee$ node dictates that exactly one of $t_2$ and $t_3$ should appear.
The topmost $\Cwedge$ node tells us the random sets derived from these $\Cvee$ nodes
coexist.

We note that and/xor trees are able to represent any finite set of possible worlds.
This can be done by listing all possible worlds, creating one $\Cwedge$ node for each world, and
using a $\Cvee$ node as the root to capture that these worlds are mutual exclusive.
Figure~\ref{eg_possibleworld} shows an example of this.

Probabilistic and/xor trees significantly generalize {\em x-tuples}~\cite{sarma:icde06,conf/icde/YiLSK08}, block-independent disjoint tuples model,
and $p$-or-sets~\cite{conf/pods/DalviS07}, and as discussed above, can represent a finite set of arbitrary possible worlds. The correlations
captured by such a tree can be represented by probabilistic c-tables~\cite{GT06_models} and provenance semirings~\cite{Pods07_semirings}.
However, that does not directly imply an efficient algorithm for ranking.
We remark that Markov or Bayesian network models are able to capture more general correlations
in a compact way~\cite{sen:vldbj09},
however, the structure of the model is more complex and probability
computations on them (inference) is typically exponential in the treewidth of the model.
The treewidth of an and/xor tree (viewing it as a Markov network) is not bounded, and hence
the techniques developed for those models can not be used to obtain polynomial time algorithms
for and/xor trees.
And/xor trees also exhibit superficial similarities to ws-trees~\cite{Koch08},
which can also capture mutual exclusivity and coexistence between tuples.
{\em We note that no prior work on ranking in probabilistic databases has considered more complex
correlations than x-tuples.}




\subsection{Ranking over Probabilistic Data: Definitions and Prior Work}
\label{sec:prior work}
\label{sec:definitions}
The interplay between probabilities and scores
complicates the semantics of ranking in probabilistic
databases. This was observed by Soliman et al.~\cite{soliman:icde07}, who first considered this
problem and  presented two definitions of \Topk\ queries in
probabilistic databases.
Several other definitions of ranking have been proposed since then. We briefly review the
ranking functions we consider in this work.

\begin{list}{--}{\leftmargin 0.15in \topsep 3pt \itemsep 3pt}
\item \emph{\bf Uncertain Top-k (\UTK)}~\cite{soliman:icde07}: Here the query returns the $k$-tuple set that appears as the top-k answer in
    most possible worlds (weighted by the probabilities of the worlds).
\item \emph{\bf Uncertain Rank-k (\URK)}~\cite{soliman:icde07}: At each rank $i$, we return the tuple with the maximum probability
    of being at the $i$'th rank in all possible worlds. In other words, \URK\ returns:\\ $\{t_{i}^{*}, i=1,2,..,\rmk
    \}$, where $t_{i}^{*}=argmax_t(\Prob(r(t)=i))$.
    Note that, under these semantics, the same tuple may be ranked at multiple positions. In our experiments, we use a
    slightly modified version that enforces distinct tuples in the answer (by not choosing a tuple at a position if it
    is already chosen at a higher position).
\item \emph{\bf Probabilistic Threshold Top-k (\PT)}~\cite{conf/sigmod/HuaPZL08}\footnote{This is quite similar to the Global Top-k semantics~\cite{conf/dbrank/ZhangC08}.}:
The original definition of a probabilistic
threshold query asks for all tuples with probability of being
in top-$h$ answer larger than a pre-specified threshold,
i.e., all tuples $t$ such that $\Prob(r(t) \leq h) > threshold$.
For consistency with other ranking functions,
we slightly modify the definition and instead ask for the $\rmk$ tuples with the largest $\Prob(r(t) \leq h)$ values.
\item \emph{\bf Expected Ranks (\ERK)}~\cite{Cormode09}:
The tuples are ranked in the
increasing order by the {\em expected} value of their ranks across the possible worlds, i.e., by:\\[2pt] \centerline{$\sum_{pw \in PW} \Prob(pw) r_{pw}(t)$,}\\[2pt] where $r_{pw}(t)$ is defined to be $|pw|$ if
$t \notin pw$.
\item \emph{\bf Expected Score (\ES)}: Another natural ranking function,
also considered by~\cite{Cormode09}, is simply
to rank the tuples by their expected score, $\Prob(t) \score(t)$.
\item {\bf $k$-selection Query~\cite{conf/dasfaa/LiuYXTL10}:} A {\em $k$-selection} query returns the set of $k$ tuples, such that the
    expected score of the best available tuple across the possible worlds is maximized.
\item \emph{\bf Consensus Top-k (\CON)}: This is a semantics for top-$k$ queries 
    developed under the framework of {\em consensus answers} in probabilistic databases~\cite{pods09_LD}.
    We defer its definition till Section~\ref{sec:contopk} where we discuss in detail its relationship with the \PRF\ function proposed in this article.
\end{list}


\topic{Normalized Kendall Distance}
To compare different ranking functions or criteria, we need a distance measure to evaluate the closeness of two top-k answers.
We use the prevalent {\em Kendall tau} distance defined for comparing \Topk\ answers for this purpose~\cite{fagin:soda03}.
It is also called {\em Kemeny distance} in the literature and is considered to
have many advantages over other distance metrics~\cite{conf/www/rankaggregation}.
Let $\mathcal{R}_1$ and $\mathcal{R}_{2}$ denote two full ranked lists, and
let $\mathcal{K}_1$ and $\mathcal{K}_{2}$ denote the top-k ranked tuples
in $\mathcal{R}_1$ and $\mathcal{R}_{2}$ respectively.
Then {\em Kendall tau distance} between $\mathcal{K}_1$ and $\mathcal{K}_{2}$ is
defined to be:\\[4pt] \centerline{$\dist(\mathcal{K}_1,\mathcal{K}_{2})=\sum_{(i, j) \in P(\mathcal{K}_1,\mathcal{K}_{2})} \hat{K}(i,j)$}, \\[4pt]
where $P(\mathcal{K}_1,\mathcal{K}_{2})$ is
the set of all unordered pairs of $\mathcal{K}_1\cup\mathcal{K}_{2}$; $\hat{K}(i,j)=1$
if it can be inferred from  $\mathcal{K}_1$ and $\mathcal{K}_{2}$
that $i$ and $j$ appear in opposite order in the two full ranked lists $\mathcal{R}_1$ and $\mathcal{R}_{2}$,
otherwise $\hat{K}(i,j)=0$.  Intuitively the Kendall distance measures the number of inversions or flips between
the two rankings.  For ease of comparison, we divide the Kendall distance by $\rmk^2$ to obtain
{\em normalized Kendall} distance, which always lies in $[0,1]$.

A higher value of the Kendall distance indicates a larger disagreement between the two \Topk\ lists.
It is easy to see that if the Kendall distance
between two top-k answers is $\delta$, then the two answers must share at least $1 - \sqrt \delta$
fraction of tuples (so if the distance is 0.09, then the \Topk\ answers share
at least 70\%, and typically 90\% or more tuples).
The distance is 0 if two \Topk\ answers are identical and 1 if they are disjoint.

\begin{table}[h]
{\footnotesize
\renewcommand{\arraystretch}{1.1}
\centering
\begin{tabular}{|c|c|c|c|c|c|}
  \hline
  & \ES & \PTPARAM{100} & \URK & \ERK & \UTK \\
\hline
\ES   & -- & 0.1241 & 0.3027   & 0.7992 & 0.2760\\
\hline
\PTPARAM{100} & 0.1241 & -- & 0.3324   & 0.9290 & 0.3674 \\
\hline
\URK  & 0.3027 & 0.3324 & --   & 0.9293 & 0.2046 \\
\hline
\ERK  & 0.7992 & 0.9290 & 0.9293   & -- & 0.9456\\
\hline
\UTK  & 0.2760 & 0.3674 & 0.2046   & 0.9456 & --\\
\hline
\end{tabular}
\vspace{2pt}

\centerline{\normalsize \bf IIP-100,000 ($\rmk=100$)}
\smallskip

\begin{tabular}{|c|c|c|c|c|c|}
  \hline
  & \ES & \PTPARAM{100} & \URK & \ERK & \UTK\\
\hline
\ES   & -- &  0.8642 & 0.8902 & 0.0044 & 0.9258\\
\hline
\PTPARAM{100} & 0.8642 &  -- & 0.3950 & 0.8647 & 0.5791 \\
\hline
\URK  & 0.8902 &  0.3950 & -- & 0.8907 & 0.3160 \\
\hline
\ERK  & 0.0044 &  0.8647 & 0.8907 & --  & 0.9263\\
\hline
\UTK  & 0.9258 & 0.5791 & 0.3160 & 0.9263 & --\\
\hline
\end{tabular}
\vspace{2pt}

\centerline{\normalsize \bf Syn-IND Dataset with 100,000 tuples ($\rmk=100$)}
\caption{Normalized Kendall distance between top-k answers according to various ranking functions for two datasets}
\label{table:compareind1}
\label{table:compareind2}
}
\end{table}

\topic{Comparing Ranking Functions}
We compared the top-100 answers returned by five of the ranking functions with each other using the normalized Kendall distance,
for two datasets with 100,000 independent tuples each (see Section \ref{sec:experiments} for a description of the datasets).
Table \ref{table:compareind1} shows the results of this experiment. 
As we can see, the five ranking functions return wildly different top-k answers for the two datasets, with
no obvious trends.
For the first dataset, \ERK\ behaves very differently from all other functions, whereas for the
second dataset, \ERK\ happens to be quite close to \ES. 
However both of them deviate largely from \UTK, \PT, and \URK.
The behavior of \ES\ is very sensitive to the dataset, especially the score distribution: it is close to \PT\ and \URK\ for the first dataset,
but far away from all of them in the second dataset (by looking into the results,
it shares less than 15 tuples with the Top-$100$ answers of the others).
We observed similar behavior for other datasets, and for datasets with correlations.


This simple experiment illustrates the issues with ranking in probabilistic databases -- although several of these
definitions seem natural, the wildly different answers they return indicate that none of them could be the ``right''
definition.

We also observe that in large datasets,
\ERK\ tends to give very high priority to a tuple with a high probability
even if it has a low score. 
In our synthetic dataset Syn-IND-100,000 with expected size $\approx 50000$,
$t_2$ (the tuple with 2nd highest score) has probability approximately 0.98 and
$t_{1000}$ (the tuple with 1000th highest score) has probability $0.99$.
The expected ranks of $t_2$ and $t_{1000}$ are approximately 10000 and 6000 respectively,
and hence $t_{1000}$ is ranked above $t_2$ even though $t_{1000}$
is only slightly more probable. 

As mentioned above, the original \URK\ function may return the same tuple at different ranks (also
observed by the authors~\cite{soliman:icde07}), which is usually undesirable.
This problem becomes even severe when the dataset and $k$ are both large.
For example, in RD-100,000, the same tuple is ranked at positions 67895 to 100000.
In the table, we show a slightly modified version of \URK\ to enforce distinct tuples in the answer.


\subsection{Parameterized Ranking Functions}
Ranking in uncertain databases is inherently a multi-criteria optimization problem, and it is not
always clear how to rank two tuples that dominate each other along different axes.
Consider a database
with two tuples $t_1$ (score = 100, $\Prob(t_1) = 0.5$), and $t_2$ (score = 50, $\Prob(t_2) = 1.0$). Even in
this simple case, it is not clear whether to rank $t_1$ above $t_2$ or vice versa. This is an instance of the
classic risk-reward trade-off, and the choice between these two options largely depends on the application domain and/or user preferences.


We propose to follow the traditional approach to dealing with such tradeoffs, by identifying a set of
{\em features}, by defining a parameterized ranking function over these features, and by learning the parameters (weights)
themselves using user preferences~\cite{Herbrich98,Joachims02,RankNet,Dekel04}.
To achieve this, we propose a family of ranking functions, parameterized by one or more parameters, and
design algorithms to efficiently find the top-k answer according to any ranking function from these families. Our general ranking
function, \PRF, directly subsumes some of the previously
proposed ranking functions, and can also be used to approximate other ranking functions.
Moreover, the parameters can be learned from user preferences, which allows us to adapt to
different scenarios and different application domains.

\topic{Features}
Although it is tempting to use the tuple probability and the tuple score as the features,
a ranking function based on just those two will be highly sensitive to the actual values of the scores; further, such a ranking function will
be insensitive to the correlations in the database, and hence cannot capture the rich interactions between
ranking and possible worlds.


Instead we propose to use the positional probabilities as the features: for each tuple $t$, we have $n$
features, \\[1pt] \centerline{$\Prob(r(t) = i), i = 1, \cdots, n$,}\\[1pt] where $n$ is the number of tuples
in the database. This set of features succinctly captures the possible worlds. Further,
correlations among tuples, if any, are naturally accounted for when computing the features.
We note that in most cases, we do not explicitly compute all the features, and instead design
algorithms that can directly compute the value of the overall ranking function.

\begin{table}
\center
\begin{tabular}{|c|c|}
\hline
$\Prob(\rk(t_i)=j)$ & Positional prob. of $t_i$ being ranked at position $j$ \\
\hline
$\Prob(\rk(t_i))$ & Rank distribution of $t_i$ \\
\hline
\PRF & Parameterized ranking function \\
& $\rank_{\omega}(t)=\sum_{i>0}\omega(t,i) \Prob(\rk(t)=i)$ \\
\hline
\PRFs($h$) & Special case of \PRF: $\omega(t, i) = w_i$, $w_i = 0, \forall i > h$ \\
\hline
\PRFe($\alpha$) & Special case of \PRFs: $w_i = \alpha^i, \alpha \in \mathbb{C}$ \\
\hline
\PRFl& Special case of \PRFs: $w_i = -i$ \\
\hline
$\delta(p)$ &  Delta function: $\delta(p) = 1$ if $p$ is true, $\delta(p) = 0$ o.w. \\
\hline
\end{tabular}
\caption{Notation}
\label{table:notation}
\end{table}

\label{subsec_newranking}
\topic{Ranking Functions}
Next we define a general ranking function which allows exploring the trade-offs discussed above.
\begin{definition} \label{PRF}
Let $\omega: T\times \mathbb{N}\rightarrow \mathbb{C}$ be a {\em weight function},
that maps a tuple-rank pair to a complex number.
The {\em parameterized ranking function (\PRF)}, $\rank_{\omega}: T\rightarrow \mathbb{C}$
in its most general form is defined to be:
\begin{eqnarray*}
\rank_{\omega}(t)&=&\sum_{pw:t\in pw} \omega(t,r_{pw}(t))\cdot \Prob(pw) \\
                 &=&\sum_{pw:t\in pw}\sum_{i>0}\omega(t,i)\Prob(pw\wedge r_{pw}(t)=i) \\
                 &=&\sum_{i>0}\omega(t,i)\cdot \Prob(r(t)=i).
\end{eqnarray*}
A \Topk\ query returns $\rmk$ tuples with the highest $|\rank_{\omega}|$ values.
\end{definition}

In most cases, $\omega$ is a real positive function and
we just need to find the $\rmk$ tuples with highest $\rank_{\omega}$ values.
However we allow $\omega$ to be a complex function in order to approximate other functions efficiently (see Section \ref{subsec_approx}).
Depending on the actual function $\omega$, we get different ranking functions with diverse behaviors.
Before discussing the relationship to prior ranking functions, we define two special cases.

\topic{\PRFsPARAM{h}} One important class of ranking functions is
when $\omega(t,i) = w_i$ (i.e., independent of $t$) and $w_i = 0\ \forall i>h$
for some positive integer $h$ (typically $h \ll n$).  This forms one of prevalent
classes of ranking functions used in domains such as information retrieval and machine learning,
with the weights typically learned from user preferences~\cite{Herbrich98,Joachims02,RankNet,Dekel04}.
Also, the weight function $\omega(i)=\frac{\ln 2}{\ln (i+1)}$ (called {\em discount factor}) is often used in the context of ranking documents
in information retrieval~\cite{journals/tis/jarvelin02}.

\topic{\PRFe($\alpha)$} This is a special case of \PRFsPARAM{h} where
$w_i = \omega(i) = \alpha^i$, where $\alpha$ is a constant and may be a real or a
complex number. Here $h = n$ (no weights are 0). Typically we expect $|\alpha| \le 1$, otherwise we have the counterintuitive
behavior that tuples with lower scores are preferred.

\smallskip
\smallskip
\noindent{\PRFs} and \PRFe\ form the two parameterized ranking functions that we propose in this work.
Although \PRFs\ is the more natural ranking function and has been used elsewhere,
\PRFe\ is more suitable for ranking in probabilistic
databases for various reasons. First, the features as we have defined above are not completely arbitrary, and
the features $\Prob(r(t) = i)$ for small $i$ are clearly more important than the ones for large $i$.
Hence in most cases we would like the weight function, $\omega(i)$, to be monotonically non-increasing.
\PRFe\ naturally captures this behavior (as long as $|\alpha| \le 1$).
More importantly, we can compute the \PRFe\ function in $O(n\log(n))$ time
($O(n)$ time if the dataset is pre-sorted by score) even for datasets with low degrees of correlations
(i.e., modeled by and/xor trees with low heights).
This makes it significantly more attractive for ranking over large datasets.

Furthermore, ranking by \PRFe($\alpha$), with suitably chosen $\alpha$,
can approximate rankings by many other functions reasonably well
even with only real $\alpha$.
Finally, a linear combination of exponential functions, with complex bases, is known to be
very expressive
in representing other functions~\cite{Beylkin05}. We make use of this fact to approximate
many ranking functions by linear combinations of a small number of
\PRFe\ functions, thus significantly speeding up the running time (Section \ref{subsec_approx}).

\topic{Relationship to other ranking functions}
We illustrate some of the choices of weight function, and relate them to prior ranking functions\footnote{The definition of the \UTK\ introduced in \cite{soliman:icde07}
requires the retrieved $\rmk$ tuples belongs to a valid possible world. However, it is not required
in our definition, and hence it is not possible to simulate \UTK\ using \PRF.}.
We omit the subscript $\omega$ if the context is clear.
Let $\delta(p)$ denote a delta function where $p$ is a boolean predicate: $\delta(p) = 1$ if $p = true$, and $\delta(p) = 0$ otherwise.

\begin{list}{--}{\leftmargin 0.15in \topsep 3pt \itemsep 3pt}

\item {\bf Ranking by probabilities:} If $\omega(t,i)=1$, the result is the set of $\rmk$ tuples with the highest
    probabilities~\cite{conf/icde/ReDS07}.

\item {\bf Expected Score:} By setting $\omega(t,i)=\score(t)$, we get the \ES:
\begin{align*} \rank(t)=\sum_{pw:t\in pw}\score(t)\Prob(pw)=\score(t)\Prob(t)
=\Exp[\score(t)]
\end{align*}

\eat{
\item \underline{\PRFsPARAM{h}:}
    One important class of ranking functions is
    when $\omega(t,i) = w_i$ (i.e., independent of $t$) and $w_i = 0 \forall i>h$ for some positive integer $h$.
    This forms one of prevalent classes of ranking functions used in
    domains such as information retrieval and machine learning,
    with the weights typically learned from user preferences~\cite{Herbrich98,Joachims02,RankNet,Dekel04}.
    Also, the weight function $\omega(i)=\frac{\ln 2}{\ln (i+1)}$ (called {\em discount factor}) is often used in the context of ranking documents
    in information retrieval~\cite{journals/tis/jarvelin02}.
    }

\item {\bf Probabilistic Threshold Top-k (\PTPARAM{h}):} If we choose
    $\omega(i) = \delta(i \le h)$, i.e.,
    $\omega(i) = 1$ for $i \le h$, and $= 0$ otherwise, then we have
    exactly the answer for \PTPARAM{h}.


\item {\bf Uncertain Rank-k (\URK):} Let $\omega_j(i) = \delta(i = j)$,
    for some $1\leq j\leq \rmk$. We can see the tuple with largest $\rank_{\omega_j}$ value is
    the rank-$j$ answer in \URK\ query~\cite{soliman:icde07}.
    This allows us to compute the \URK\ answer by evaluating $\rank_{\omega_j}(t)$ for all $t\in T$ and $j=1, \dots, \rmk$.

\item {\bf Expected ranks (\ERK)}: Let \PRFl\ (\PRF\ linear) be another special case of the \PRFs\ function,
    where $w_i = \omega(i) = -i$. The \PRFl\ function bears a close similarity to the notion
    of {\em expected ranks}.
    Recall that the expected rank of a tuple $t$ is defined to be:
    \[
    \Exp[r_{pw}(t)] = \sum_{pw\in PW} \Prob(pw) r_{pw}(t)
    \]
    where $r_{pw}(t) = |pw|$ if $t_i \notin pw$. Let $C$ denote the expected size of a possible world.
    It is easy to see that: $C =\sum_{i=1}^n p_i$ due to linearity of expectation.
    Then the expected rank of $t$ can be seen to consist of two parts:
    \begin{list}{--}{\leftmargin 0.19in \topsep 2pt \itemsep 2pt}
    \item[(1)] the contribution of possible worlds where $t$ exists:
        \[ er_1(t) = \sum_{i > 0} i \times \Prob(r(t) = i) = -\rank(t) \]
        where $\rank(t)$ is the \PRFl\ value of tuple $t$.\footnote{Note that, in the expected rank approach, we pick the $k$ tuples
        with the {\em lowest} expected rank, but in our approach, we choose the tuples with the {\em highest} PRF function values, hence the negation.}
    \item[(2)] the contribution of worlds where $t$ does not exist:
        \begin{eqnarray*}
    er_2(t) &=& \sum_{pw:t\notin pw} \Prob(pw) |pw|  \\
    &=&  (1 - p(t))(\sum_{t_i \ne t} \Prob(t_i \mid t\ \text{does not exist}))
        \end{eqnarray*}
\end{list}

If the tuples are independent of each other, then we have:
\[
\sum_{t_i \ne t} \Prob(t_i \mid t\ \text{does\ not\ exist}) = (C - p(t))
\]
Thus, the expected ranks can be computed in the same time as \PRFl\ in tuple-independent datasets.
This term can also be computed efficiently in many other cases, including in datasets where only
mutual exclusion correlations are permitted.
If the correlations are represented using a probabilistic and/xor tree (see Section \ref{subsec_andortree})
or a low-treewidth
graphical model (see Section \ref{sec:correlations}), then we can compute this term efficiently as well,
thus generalizing the prior algorithms for computing expected ranks.



\item {\bf $k$-selection Query~\cite{conf/dasfaa/LiuYXTL10}:} It is easy to see that a $k$-selection query is equivalent to setting: 
    $\omega(t, i) = \delta(i = 1) \score(t)$.
\end{list}

As we can see, many different ranking functions can be seen as special cases of the general
\PRF\ ranking function, supporting our claim that \PRF\ can effectively unify these different
approaches to ranking uncertain datasets.

\section{Ranking Algorithms}
\label{sec:ranking algorithms}
\label{sec:algorithms}
We next present an algorithm for efficiently ranking according to a \PRF\ function. We first present
the basic idea behind our algorithm assuming mutual independence, and then consider correlated tuples
with correlations represented using an and/xor tree. We then present a very efficient algorithm for
ranking using a \PRFe\ function, and then briefly discuss how to handle attribute uncertainty.

\subsection{Assuming Tuple Independence}
\label{subsec_PRF_ind}

First we show how the \PRF\ function can be computed in $O(n^2)$ time for a general weight function $\omega$,
and for a given set of tuples $T = \{t_1, \ldots, t_n\}$.
In all our algorithms, we assume that $\omega(t, i)$ can be computed in $O(1)$ time.

Clearly it is sufficient
to compute $\Prob(r(t) = j)$ for any tuple $t$ and $1 \leq j \leq n$ in $O(n^2)$ time. Given these
values, we can directly compute the values of $\rank(t)$ in $O(n^2)$ time. (Later, we will present
several algorithms which run in $O(n)$ or $O(n\log(n))$ time which combine these two steps
for some special $\omega$ functions).

We first sort the tuples in a non-increasing order
by their scores (which are assumed to be deterministic); assume $t_1, \dots, t_n$ indicates this sorted order.
Suppose now we want to compute $\Prob(r(t_i)=j)$.
Let
$T_i=\{t_1,t_2,\ldots,t_i\}$ and
$\sigma_i$ be an indicator variable that takes value $1$ if $t_i$ is present in a possible world, and $0$ otherwise.
Further, let $\sigma = \langle \sigma_1, \dots, \sigma_n \rangle$ denote a vector
containing all the indicator variables. Then, we can write $\Prob(r(t_i) = j)$ as follows:
\begin{eqnarray*}
\Prob(r(t_i)=j)
&=&
\Prob(t_i)\sum_{pw:|pw\cap T_{i-1}|=j-1}\Prob(pw) \\
&=&
\Prob(t_i)
\sum_{\sigma: \sum\limits_{l=1}^{i-1}\sigma_l=j-1}
\prod_{l<i:\sigma_l=1}\Prob(t_l)
\prod_{l<i:\sigma_l=0}(1-\Prob(t_l))
\end{eqnarray*}

The first equality says that tuple $t_i$ ranks at the $j$th position if and only if
$t_i$ {\em and} exactly $j-1$ tuples from $T_{i-1}$ are present in the possible world. The second equality
is obtained by rewriting the sum to be over the indicator vector (each assignment to the indicator
vector corresponds to a possible world), and by exploiting the fact that
the tuples are independent of each other. The naive method to evaluate the above formula by explicitly listing all possible worlds needs
exponential time.  Now, we present a polynomial time algorithm based on generating functions.

\noindent{Consider} the following generating function over $x$:\\[2pt] \centerline{$\calF(x)=\prod_{i=1}^n(a_i+b_ix)$}\\[2pt]
The coefficient of $x^k$ in $\calF(x)$ is:\\[2pt]
\centerline{$\sum_{|\mathbf{\beta}| =k}\prod_{i:\beta_i=0} a_i\prod_{i:\beta_i=1} b_i$} \\[2pt] where
$\mathbf{\beta}= \langle \beta_1,\ldots,\beta_n \rangle$ is a boolean vector,
and $|\mathbf{\beta}|$ denotes the number of $1$'s in $\mathbf{\beta}$. 
Now consider the following generating function:
\begin{eqnarray*}
\calF^i(x)=\biggl(\prod_{t\in T_{i-1}} \bigl(1-\Prob(t)+\Prob(t)\cdot x \bigr)\biggr)\Prob(t_i)\cdot x
        =\sum_{j\geq 0}c_j x^j.
\end{eqnarray*}

\noindent{We} can see that the coefficient $c_j$ of $x^j$ in the expansion of $\calF^i$ is
exactly the probability that $t_i$ is at rank $j$, i.e., $c_j = \Prob(r(t_i)=j)$.
We note $\calF^i$ contains at most $i+1$ nonzero terms. We observe this both from the form of $\calF^i$ above,
and also from the fact that $\Prob(r(t_i)=j) = 0$ if $j > i$.
Hence, we can expand $\calF^i$ to compute the coefficients in $O(i^2)$ time.
This allows us to compute
$\Prob(r(t_i) = j)$ for $t_i$ in $O(i^2)$ time; $\rank(t_i)$, in turn, can be written as:
\begin{equation}
\label{eqn_rank_gene}
\rank(t_i)=\sum_{j}\omega(t_i,j)\cdot \Prob(r(t_i)=j)=\sum_j \omega(t_i,j)c_j
\end{equation}
which can be computed in $O(i^2)$ time.

\begin{figure*}
\centerline{\includegraphics[width=0.9\linewidth]{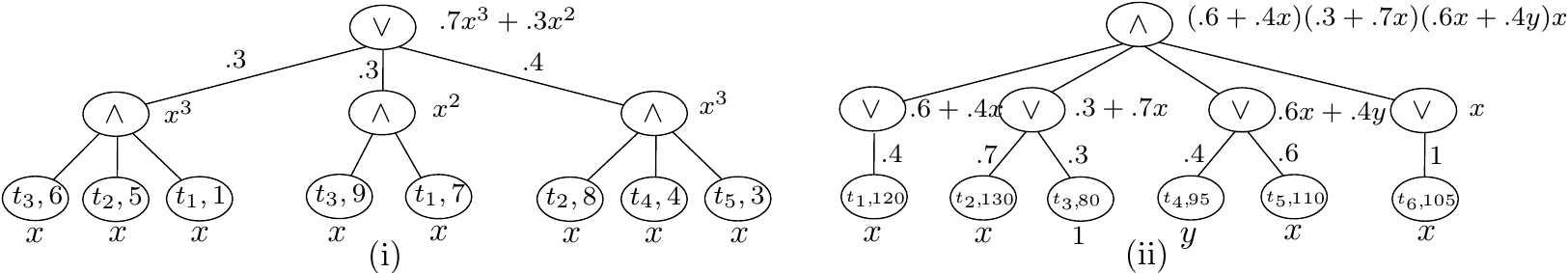}}
\caption{\PRF\ computation on and/xor trees: (i) The left figure corresponds to the
database in Figure \ref{eg_possibleworld}; the generating function obtained by assigning the same variable
$x$ to all leaves gives us the distribution over the sizes of the possible worlds.
(ii) The right figure illustrates the construction of the generating function for computing $\Prob(r(t_{4})=3)$ in the and/xor tree
in Figure \ref{fig_example}.}
\label{fig_andortree}
\end{figure*}

\eat{
\begin{algorithm}
\caption{IND-PRF-RANK($D_T$)}
\label{algo_indepentPRF}
\begin{algorithmic}
 \STATE $\calF^0(x)=1$
 \FOR{i=1 to n}
  \STATE $\calF^i(x)={\Prob(t_i) \over \Prob(t_{i-1})}\calF^{i-1}(x)(1-\Prob(t_{i-1})+\Prob(t_{i-1})x)$
  \STATE Expand $\calF^i(x)$ in the form of $\sum_j c_jx^j$
  \STATE $\rank(t_i)=\sum_{j=1}^n \omege(t_i,j)c_j$.
 \ENDFOR
 \STATE \textbf{return} $\rmk$ tuples with largest $\rank$ values.
\end{algorithmic}
\end{algorithm}
}

\begin{example}
\label{eg_independent}
Consider a relation with 3 independent tuples $t_1$, $t_2$, $t_3$ (already sorted according to the score function) with
existence probabilities $0.5, 0.6, 0.4$, respectively.
The generating function for $t_3$ is: \\[3pt]
\centerline{$\calF^3(x)=(.5+.5x)(.4+.6x)(.4x)= .12 x^3+ .2 x^2+ .08 x$} \\[1pt]
This gives us:\\[3pt]
\centerline{$\Prob(r(t_3)=1)=.08,\Prob(r(t_3)=2)=.2,\Prob(r(t_3)=3)=.12$}
\end{example}

If we expand each $\calF^i$ for $1\leq i\leq n$ from scratch, we need $O(n^2)$ time for each $\calF^i$ and
$O(n^3)$ time in total.
However, the expansion of $\calF^i$ can be obtained from the expansion of $\calF^{i-1}$ in $O(i)$ time
by observing that:
\begin{equation}
\label{eqn_rank_gene_recursion}
\calF^i(x)={\Prob(t_i) \over \Prob(t_{i-1})}\calF^{i-1}(x)\Bigl(1-\Prob(t_{i-1})+\Prob(t_{i-1})x\Bigr)
\end{equation}
This trick gives us a $O(n^2)$ time complexity
for computing the values of the ranking function for all
tuples. See Algorithm \ref{algo_indepentPRF} for the pseudocode.
Note that $O(n^2)$ time is asymptotically optimal in general since the computation involves at least
$O(n^2)$ probabilities, namely $\Prob(r(t_i)=j)$ for all $1\leq i,j\leq n$.

\linesnumbered
\begin{algorithm}[t]
\caption{$\mathsf{IND}$-$\mathsf{PRF}$-$\mathsf{RANK}$($D_T$)}
\label{algo_indepentPRF}
$\calF^0(x)=1$\;
\For{$i=1$ to $n$} {
    $\calF^i(x)={\Prob(t_i) \over \Prob(t_{i-1})}\calF^{i-1}(x)\Bigl(1-\Prob(t_{i-1})+\Prob(t_{i-1})x\Bigr)$\;
    Expand $\calF^i(x)$ in the form of $\sum_j c_jx^j$ \;
    $\rank(t_i)=\sum_{j=1}^n \omega(t_i,j)c_j$ \;
    }
\textbf{return} $\rmk$ tuples with largest $\rank$ values\;
\end{algorithm}

For some specific $\omega$ functions, we may be able to achieve faster running time.
For \PRFsPARAM{h} functions,
we only need to expand all $\calF^i$'s up to $x^h$ term since $\omega(i)=0$ for $i>h$.
Then, the expansion from $\calF^{i-1}(x)$ to $\calF^i(x)$ only takes $O(h)$ time.
This yields an $O(n\cdot h+n\log(n))$ time algorithm.
We note the above technique also gives an $O(n\rmk+n\log(n))$ time algorithm
for answering the \URK\ \Topk\ query (all the needed probabilities can be computed in that time), thus matching the best known upper bound by Yi et al.~\cite{conf/icde/YiLSK08}
(the original algorithm in~\cite{soliman:icde07} runs in $O(n^2\rmk)$ time).

We remark that the generating function technique can be seen as a variant of dynamic programming in
some sense; however, using it explicitly in place of the obscure recursion formula gives us a much
cleaner view and allows us to generalize it to handle more complicated tuple correlations.
This also leads to an algorithm for extremely efficient evaluation of \PRFe\ functions (Section \ref{subsec_PRFe}).

\subsection{Probabilistic And/Xor Trees}
\label{subsec_andortree}

Next we generalize our algorithm to handle a correlated database where the correlations can be
captured using an {\em and/xor} tree.
In fact, many types of probability computations on and/xor trees
can be done efficiently and elegantly using generating functions.
Here we first provide a general result and then
specialize it for \PRF\ computation.

As before, let $T =\{t_1, t_2, \ldots, t_n\}$ denote the tuples sorted in a non-increasing order
of their score function, and let $T_i = \{t_1, t_2, \ldots, t_i\}$.
Let $\calT$ denote the and/xor tree.
Suppose $\calX=\{x_1, x_2, \ldots \}$ is a set of variables. Define a mapping $\pi$
which associates each leaf $l \in \calT$ with a variable $\pi(l) \in \calX$.
Let $\calT_v$ denote the subtree rooted at $v$ and let $v_1, \ldots, v_h$ be $v$'s children.
For each node $v \in \calT$, we define a generating function
$\calF_v(\calX)=\calF_{v}(x_{1},x_{2},\ldots)$ recursively: 

\begin{list}{$\bullet$}{\leftmargin 0.13in \topsep 2pt \itemsep 2pt}
\item
If $v$ is a leaf, $\calF_v(\calX)=\pi(v)$.
\item
If $v$ is a $\Cvee$ node, \\[5pt]
$
\calF_v(\calX)=(1-\sum_{l=1}^h p_{(v,v_l)})+ \sum_{l=1}^h p_{(v,v_l)}\calF_{v_l}(\calX)
$
\item
If $v$ is a $\Cwedge$ node, $\calF^i_v(\calX)=\prod_{l=1}^h \calF_{v_l}(\calX) $.
\end{list}

\smallskip
\smallskip
The generating function $\calF(\calX)$ for tree $\calT$ is the one defined above for the root.
It is easy to see, if we have a constant number of variables, the polynomial can be expanded in
the form of $\sum_{i_1,i_2,\ldots}c_{i_1,i_2\ldots}x_1^{i_1}x_2^{i_2}\ldots$ in polynomial time.

Now recall that each possible world $pw$ contains a subset of the leaves of $\calT$ (as dictated
by the $\Cvee$ and $\Cwedge$ nodes).
The following theorem characterizes the relationship between the coefficients of $\calF$ and
the probabilities we are interested in.
\begin{theorem}
\label{thm_generating}
The coefficient of the term $\prod_{j}x_j^{i_j}$ in $\calF(\calX)$ is the total probability of
the possible worlds for which, for all $j$, there are exactly $i_j$ leaves associated with variable $x_j$.
\end{theorem}

See Appendix \ref{sec:proofs} for the proof.
We first provide two simple examples to show how to use Theorem~\ref{thm_generating}
to compute the probabilities of two events related to the size of the possible world,
and then show how to use the same idea to compute $\Pr(r(t) = i)$.
\begin{example}
If we associate all leaves with the same variable $x$, the coefficient of $x^i$ is equal to $\Prob(|pw|=i)$.
The above can be used to obtain a distribution on the possible world sizes (Figure \ref{fig_andortree}(i)).
\end{example}
\begin{example}
If we associate a subset $S$ of the leaves with variable $x$, and other leaves with constant $1$,
the coefficient of $x^i$ is equal to $\Prob(|pw\cap S|=i)$.
\end{example}

\linesnotnumbered
\begin{algorithm}[h]
\caption{ $\mathsf{ANDXOR}$-$\mathsf{PRF}$-$\mathsf{RANK}$($\calT$)}
 $\pi(t_{i})\leftarrow 1 \forall i$ \{$\pi(t_{i})$ is the variable associated to leaf $t_{i}$\}\;
 \For{$i=1$ to $n$}{
   \textbf{if} {$i\ne 1$} \textbf{then} $s(t_{i-1})\leftarrow x$\;
  $\pi(t_{i})\leftarrow y$\;
  $\calF^i(x,y)= \mathsf{GENE}(\calT_i,\pi)$\;
  Expand $\calF^i(x,y)$ in the form $\sum_j c'_jx^j + (\sum_j c_jx^{j-1})y$\;
  $\rank(t_i)=\sum_{j=1}^n \omega(t_i,j)c_j$\;
 }
 \textbf{return} $\rmk$ tuples with largest $\rank$ values\;

 \textbf{Subroutine:} $\mathsf{GENE}(\calT,\pi)$\;
 $r$ is the root of tree $\calT$\;
 \eIf{$\calT$ is a singleton node}{
       \textbf{return} $\pi(r)$\;
   }{
   $\calT_i$ is the subtree rooted at $r_i$ for $r_i\in Ch(r)$\;
   $p=\sum_{r_i\in Ch(r)} p_{(r,r_i)}$\;
   \If{$r$ is a \textcircled{\small{$\Cvee$}} node}{
       \textbf{return} $1-p+\sum_{r_i\in Ch(r)} p_{(r,r_i)}\cdot \mathsf{GENE}(\calT_i,t)$\;
   }
   \If{$r$ is a $\Cwedge$ node}{
       \textbf{return} $\prod_{r_i\in Ch(r)} \mathsf{GENE}(\calT_i,t)$\;
   }
 }
\label{fig_algo_andor}
\end{algorithm}

Next we show how to compute $\Prob(r(t_{i}) = j)$ (i.e., the probability $t_{i}$ is ranked at position $j$).
Let $s$ denote the score of the tuple.
In the and/xor tree $\calT$, we associate all leaves 
with score value larger than $s$ with variable $x$,
the leaf $(t_{i},s)$ with variable $y$, and the rest of leaves with constant $1$.
Let the resulting generating function be $\calF^{i}$.
By Theorem~\ref{thm_generating}, the coefficient of $x^{j-1}y$ in the generating function $\calF^{i}$ is exactly
$\Prob(r(t_{i}) = j)$.
See Algorithm \ref{fig_algo_andor} for the pseudocode of the algorithm.


\begin{example}
We consider the database in Figure~\ref{fig_example}.
Suppose we want to compute $\Prob(r(t_{4})=3)$.
We associate variable $x$ to $t_1, t_2, t_5$ and $t_6$ since their scores are larger than $t_4$'s score.
We also associate $y$ to $t_4$ itself and $1$ to $t_3$ whose score is less $t_4$'s.
The generating function for the right hand side tree in
Figure \ref{fig_andortree} is $(.6+.4x)(.3+.7)(.4x+.6y)x=.168x^{4}+0.112x^{3}y+0.324x^{3}+0.216x^{2}y+0.108x^{2}+0.072xy$.
So we get that $\Prob(r(t_5)=3)$ is the coefficient of $x^{2}y$ which is $0.216$. From Figure \ref{fig_example},
we can also see $\Prob(r(t_5)=3)=\Prob(pw_3)+\Prob(pw_5)=.048+.168=.216$. 
\end{example}

If we expand $\calF^i_v$ for each internal node $v$ in a naive way (i.e., we do polynomial
multiplication one by one), we can show the running time is $O(n^2)$ at each
internal node, $O(n^{3})$ for each tree $\calF^{i}$ and thus $O(n^4)$ overall.
If we do divide-and-conquer at each internal node
and use the FFT-based (Fast Fourier Transformation) algorithm for the multiplication of polynomials,
the running time for each $\calF^{i}$
can be improved to $O(n^2\log^2(n))$. See Appendix~\ref{app_runningtime} for the details.
In fact, we can further improve the running time to $O(n^{2})$ for each $\calF^{i}$
and $O(n^3)$ overall.
We outline two algorithms in Appendix~\ref{app_runningtime2}.

\subsection{Computing a \PRFe\ Function}
\label{subsec_PRFe}

Next we present an $O(n\log(n))$ algorithm to evaluate a \PRFe\ function (the algorithm runs in linear time if
the dataset is pre-sorted by score). If $\omega(i)=\alpha^i$, then we observe that:
\begin{equation}
\label{eqn_prfe}
\rank(t_i)=\sum_{j=1}^n\Prob(r(t_i)=j)\alpha^j=\calF^i(\alpha)
\end{equation}
This surprisingly simple relationship suggests we don't have to expand the polynomials $\calF^i(x)$ at all;
instead we can evaluate the numerical value of $\calF^i(\alpha)$ directly.
Again, we note that the value $\calF^i(\alpha)$ can be computed from the value of
$\calF^{i-1}(\alpha)$ in $O(1)$ time using Equation (\ref{eqn_rank_gene_recursion}).
Thus, we have $O(n)$ time algorithm to compute $\rank(t_i)$ for all $1\leq i\leq n$
if the tuples are pre-sorted.

\begin{example}
\label{eg_prfe}
Consider Example \ref{eg_independent} and the $PRF^e$ function for $t_3$.
We choose $\omega(i)=.6^i$.
Then, we can see that $\calF^3(x) = (.5+.5x)(.4+.6x)(.4x)$. So,
$\rank(t_3)=\calF^3(.6)=(.5+.5\times .6)(.4+.6\times .6)(.4\times .6)=.14592.$
\end{example}

\noindent{We} can use a similar idea to speed up the computation if the tuples are correlated and the
correlations are represented using an and/xor tree.
Let $\calT_i$ be the and/xor tree where $\pi(t_{j})=x$ for $1\leq j<i$, $\pi(t_{i})=y$ and $\pi(t_{j})=1$ for $j>i$.
Suppose the generating function for $\calT_i$ is
$\calF^i(x,y)=\sum_j c'_jx^j + (\sum_j c_jx^{j-1})y$ and  $\rank(t_i)=\sum_{j=1}^n \alpha^j c_j$.
We observe an intriguing relationship between the
\PRFe\ value and the generating function:
\begin{eqnarray*}
\rank(t_i)&=&\sum_j  c_j\alpha^j=
\Bigl(\sum_j  c'_j\alpha^j + (\sum_j  c_j\alpha^{j-1})\alpha\Bigr)-
\sum_j  c'_j\alpha^j \\
&=& \calF^i(\alpha,\alpha)-\calF^i(\alpha,0).
\end{eqnarray*}
Given this, $\rank(t_i)$ can be computed in linear time by bottom up evaluation
of $\calF^i(\alpha,\alpha)$ and $\calF^i(\alpha,0)$ in $\calT^i$.
If we simply repeat it $n$ times, once for each $t_i$,
this gives us a $O(n^2)$ total running time.

By carefully sharing the intermediate results among
computations of $\rank(t_i)$, we can improve the running time to $O(n\log(n) + nd)$
where $d$ is the height of the and/xor tree.
This improved algorithm runs in
iterations. Suppose the tuples are already pre-sorted by their scores.
Initially, the label of all leaves, i.e., $\pi(t_{i})$, is $1$.
In iteration $i$, we change the label of leaf $t_{i-1}$ from $y$ to $x$
and the label of $t_{i}$ from $1$ to $y$.
The algorithm maintains the following information in each inner node $v$:
the numerical values of $\calF^i_v(\alpha,\alpha)$ and $\calF^i_v(\alpha,0)$.
The values on node $v$ need to be updated when the value of one of its children changes.
Therefore, in each iteration, the computation only happens on the two paths,
one from $t_{i-1}$ to the root and one from $t_{i}$ to the root.
Since we update at most $O(d)$ nodes for each iteration,
the running time is $O(nd)$.
Suppose we want to update the information on the path from $t_{i-1}$ to the root.
We first update the $\calF^i_v(.,.)$ values for the leaf $t_{i-1}$.
Since $\calF^{i}_{t_{i-1}}=\pi(t_{i-1})=x$, we have
$\calF^{i}_{t_{i-1}}(\alpha, \alpha)=\alpha$ and $\calF^{i}_{t_{i-1}}(\alpha, 0)=\alpha$.
We assume $v$'s child, say $u$, just had its values changed.
The updating rule for $\calF^i_v(.,.)$(both $\calF^i_v(\alpha,\alpha)$ and $\calF^i_v(\alpha,0)$) in node $v$ is as follows.
\begin{enumerate}
\item $v$ is a $\Cwedge$ node, $\calF^i_v(.,.)\leftarrow {\calF_v^{i-1}(.,.)\calF_u^{i}(.,.)/\calF_u^{i-1}(.,.)}$
\item $v$ is a $\Cvee$ node, then:
\\[3pt] $\calF^i_v(.,.)\leftarrow \calF_v^{i-1}(.,.)+p_{(v,u)}\calF_u^{i}(.,.)- p_{(v,u)}\calF_u^{i-1}(.,.)$
\end{enumerate}
The values on other nodes are not affected.
The updating rule for the path from $t_{i}$ to the root is the same except that
for the leaf $t_{i}$, we have $\calF^{i}_{t_{i}}(\alpha, \alpha)=\alpha$ and $\calF^{i}_{t_{i}}(\alpha, 0)=0$
since $\calF^{i}_{t_{i}}(x,y)=\pi(t_{i})=y$.
See Algorithm~\ref{fig_algo_andor_prfe} for the psuedo-code.

\noindent{We} note that, for the case of {\em x-tuples}, which can be represented using a two-level tree,
this gives us an $O(n\log(n))$ algorithm for ranking according to \PRFe.

\linesnotnumbered
\begin{algorithm}[h]
$\calF_{t_{i}}(\alpha, \alpha)=1,\calF_{t_{i}}(\alpha, 0)=1,\forall i$ \;
\label{fig_algo_andor_prfe}
\caption{ $\mathsf{ANDXOR}$-$\mathsf{PRF^{e}}$-$\mathsf{RANK}$($\calT$)}
 \For{$i=1$ to $n$}{
  \If{$i\ne 1$}{	
	 $\calF_{t_{i-1}}(\alpha, \alpha)=\alpha,\calF_{t_{i-1}}(\alpha, 0)=\alpha$ \;
 	$\mathsf{UPDATE}(\calT, t_{i-1})$\;
  }
 $\calF_{t_{i}}(\alpha, \alpha)=\alpha,\calF_{t_{i}}(\alpha, 0)=0$ \;
 $\mathsf{UPDATE}(\calT, t_{i})$\;
 $\rank(t_i)= \calF_{r}(\alpha,\alpha)-\calF_{r}(\alpha,0)$\;
 }
 \textbf{return} $\rmk$ tuples with largest $\rank$ values\;
  \textbf{Subroutine:} $\mathsf{UPDATE}(\calT, v)$\;
  \While{$v$ is not the root}
  {
	$u\leftarrow v$\;
  	$v\leftarrow parent(v)$\;
	\If{ $v$ is a $\Cwedge$ node}
	{
	 $\calF_v(.,.)\leftarrow {\calF_v(.,.)\calF_u^{i}(.,.)/\calF_u(.,.)}$\;
	 }
	\If{$v$ is a $\Cvee$ node}
	{
	$\calF_v(.,.)\leftarrow \calF_v(.,.)+p_{(v,u)}\calF_u(.,.)- p_{(v,u)}\calF_u(.,.)$\;
	}
  }
\label{algo_andorPRFtwo}
\end{algorithm}

\begin{table*}[t]
\center
\begin{tabular}{|c|c|c|c|}
  \hline
  & \PRF\ & \PRFs($h$)\ & \PRFe \\
  \hline
  Independent tuples  & $O(n^2)$  & $O(nh+n\log(n))$\;  & $O(n\log(n))$\\
  \hline
  And/Xor tree  (height=$d$) & $O(n^3)$ or $O(n^2\log^2(n) d)$ & $O(n^3)$ or $O(n^2\log^2 (n)d)$ & $O(nd+n\log(n)$\\
  \hline
  And/Xor tree      & $O(n^3)$ & $O(n^3)$ & $O(\sum_{i}d_i+n\log(n))$\\
  \hline
\end{tabular}
\caption{Summary of the running times. $n$ is the number of tuples.
$d_i$ is the depth of tuple $t_i$ in the and/xor tree.
}
\label{table:runningtime}
\end{table*}

\subsection{Attribute Uncertainty or Uncertain Scores}
\label{sec:attribute_uncertainty}
We briefly describe how we can do ranking over tuples with discrete attribute uncertainty where
the uncertain attributes are part of the tuple scoring function (if the uncertain attributes do
not affect the tuple score, then they can be ignored for the ranking purposes). More generally,
this approach can handle the case when there is a discrete probability distribution
over the score of the tuple.

Assume $\sum_{j}p_{i,j}\leq 1$ for all $i$.
The score $\score_i$ of tuple $t_i$ takes value $v_{i,j}$ with probability $p_{i,j}$
and $t_i$ does not appear in the database with probability $1-\sum_j p_{i,j}$.
It is easy to see the \PRF\ value of $t_i$ is
\begin{align*}
\rank(t_i)&=\sum_{k>0}\omega(t_i,k)\Prob(r(t_i)=k)\\
&=\sum_{k>0}\omega(t_i,k)\sum_{j}\Prob(r(t_i)=k \wedge \score_i=v_{i,j})\\
&=\sum_{j}\Bigl(\sum_{k>0}\omega(t_i,k)\Prob(r(t_i)=k \wedge \score_i=v_{i,j})\Bigr)
\end{align*}
The algorithm works by treating the alternatives of the tuples (with a separate alternative
for each different possible score for the tuple) as different tuples.
In other words, we create a new tuple $t_{i,j}$ for each $v_{i,j}$ value.
$t_{i,j}$ has existence probability $p_{i,j}$.
Then, we add an {\em xor} constraint over the alternatives $\{t_{i,j}\}_j$ of each tuple $t_i$.
We can then use the algorithm for the probabilistic and/xor
tree model to find the values of the \PRF\ function for each $t_{i,j}$ separately.
Note that $\Prob(r(t_i)=k \wedge \score_i=v_{i,j})$ is exactly the probability that $r(t_{i,j})=k$
in the and/xor tree. Thus, by the above equation, we have that
$\rank(t_{i,j})=\sum_{k>0}\omega(t_i,k)\Prob(r(t_i)=k \wedge \score_i=v_{i,j})$ and
$
\rank(t_i)=\sum_{j}\rank(t_{i,j}).
$
Therefore, in a final step, we calculate the $\rank$ score for each original tuple $t_i$
by adding the $\rank$ scores of its alternatives $\{t_{i,j}\}_j$.
If the original tuples were independent, the complexity
of this algorithm is $O(n^2)$ for computing the \PRF\ function, and $O(n \log(n))$ for computing
the \PRFe\ function where $n$ is the size of the input, i.e., the total number of different possible scores.


\eat{
\begin{table*}[t]
\center
\begin{tabular}{|c|c|c|c|}
  \hline
  & \PRF\ & \PRFs(h)\ & \PRFe \\
  \hline
  Independent  & $O(n^2)$  & $O(nh)$\; ($\star$) & $O(n\log(n))$\\
  \hline
  And/Xor tree  & $O(n^3)$ or & $O(n^3)$ or & \multirow{2}{*}{$O(nd)$\;($\star$)}\\
  (height=$d$)  & $O(n^2\log^2(n) d) $ & $O(n^2\log^2 (n)d)$ & \\
  \hline
  And/Xor tree      & $O(n^3)$ & $O(n^3)$ & $O(n^2)$\\
  \hline
\end{tabular}
\caption{Summary of the running times. ($\star$):
There is an additive $O(n\log(n))$ term if the dataset was not pre-sorted by their scores.
}
\label{table:runningtime}
\end{table*}
}


\subsection{Summary}
\label{subsec_summary_time}
We summarize the complexities of the algorithms for different models
in Table~\ref{table:runningtime}.
Now, we explain some entries in the table
which has not been discussed.
The first is the \PRF\ computation over an and/xor tree with height $d$.
We have two choices here. One is just to use
the algorithm for arbitrary and/xor trees,
i.e., to use the algorithm in Appendix~\ref{app_runningtime2} to
expand $\calF^i(x,y)$ for each $i$, which runs in $O(n^2)$ time.
The overall running time is $O(n^3)$.
The other one is to use the divide-and-conquer algorithm in Appendix~\ref{app_runningtime}
to expand the polynomial for each $\Cwedge$ node in $\calF^i(x,y)$.
We can easily see that expanding nodes for each level of the tree requires
at most $O(n\log^2(n))$ time. Therefore, the running time for expanding $\calF^i(x,y)$
is at most $O(n\log^2(n)d)$ and the overall running time is $O(n^2\log^2(n)d)$
which is much better than $O(n^3)$ if $d\ll n$.
For \PRFs($h$) computation over and/xor trees, we do not know how to achieve
a better bound as in the tuple-independent datasets.
We leave it as an interesting open problem.

For \PRFe\ computation on and/xor trees,
we use $\mathsf{ANDXOR}$-$\mathsf{PRF^{e}}$-$\mathsf{RANK}$.
Now, the procedure $\mathsf{UPDATE}(\calT, t_{i})$ runs in $O(d_i)$ time
where $d_i$ is the depth of tuple $t_i$ in the and/xor tree, i.e.,
the length of path from the root to $t_i$. Therefore, the total running time is $O(\sum_i d_i+n\log(n))$.
If the height of the and/xor tree is bounded by $d$, the running time is simply
$O(nd+n\log(n))$.


\section{Approximating and Learning Ranking Functions}
\label{sec:approximating and learning}
In this section, we discuss how to choose the \PRF\ functions and their parameters.
Depending on the application domain and the scenarios, there are two approaches to this:
\begin{mylist}
\item If we know the ranking function we would like to use (say \PT), then we can either simulate or approximate it
using appropriate \PRF\ functions.
\item If we are instead provided user preferences data, we can learn the parameters from them.
\end{mylist}

Clearly, we would prefer to use a \PRFe\ function, if possible, since it admits highly efficient ranking algorithms.
For this purpose, we begin with presenting an algorithm to find an approximation to an arbitrary \PRFs\ function using a linear combination of \PRFe\ functions.
We then discuss how to learn a \PRFs\ function from user preferences, and finally present an algorithm for learning a single \PRFe\ function.

\subsection{Approximating \PRFs\ using \PRFe\ Functions}
\label{subsec_approx}
\label{sec:approximating}

A linear combination of complex exponential functions is
known to be very expressive, and can approximate many other
functions very well~\cite{Beylkin05}. Specifically, given a
\PRFs\ function, if we can write $\omega(i)$ as: $\omega(i) \approx \sum_{l=1}^L
u_l \alpha_l^i$, then we have that:
$$\rank(t)=\sum_i \omega(i)\Prob(r(t)=i) \approx \sum_{l=1}^L u_l \left(\sum_i \alpha_l^i \Prob(r(t)=i) \right) $$
This reduces the computation of $\rank(t)$ to $L$ individual \PRFe\
function computations, each of which only takes linear time. This gives us
an $O(n\log(n)+nL)$ time algorithm for approximately ranking using \PRFs\
function for independent tuples (as opposed to $O(n^2)$ for exact ranking).

Several techniques have been proposed for finding such approximations
using complex exponentials~\cite{prony,Beylkin05}. Those
techniques are however computationally inefficient, involving
computation of the inverses of large matrices and the roots of polynomials
of high orders.


In this section, we present a clean and efficient algorithm, based on Discrete Fourier Transforms (DFT),
for approximating a 
function $\omega(i)$, that approaches zero for large values of $i$
(in other words, $\omega(i) \ge \omega(i+1) \forall i, \omega(i) = 0, i>h$).
As we noted earlier, this captures the typical behavior of the $\omega(i)$ function.
An example of such a function is the step function ($\omega(i) = 1 \forall i \le h, = 0 \forall i > h$)
which corresponds
to the ranking function \PT. At a high level, our algorithm starts with a DFT approximation of $\omega(i)$ and
then adapts it by adding several damping, scaling and shifting factors.



\eat{A common practice to approximate a function by linear combination of
a few complex exponentials is first to decompose the function to
possibly a large number exponentials (usually, this number is
proportion to the domain size) and then take the first few
exponentials with largest magnitudes.}
Discrete Fourier transformation (DFT) is a well known technique
for representing a function as a linear combination of complex exponentials
(also called {\em frequency domain representation}). More specifically, a
discrete function $\omega(i)$
defined on a finite domain $[0,N-1]$ can be decomposed into exactly $N$
exponentials as:
\begin{equation}
\label{eqn} \omega(i)=\frac{1}{N}\sum_{k=0}^{N-1}
\psi(k) e^{\frac{2\pi \uimag}{N} k i} \quad \quad i = 0,\ldots,N-1.
\end{equation}
where $\uimag$ is the imaginary unit and
$\psi(0),\cdots,\psi(N-1)$ denotes the DFT transform of $\omega(0), \cdots$, $\omega(N-1)$.
If we want to approximate $\omega$ by fewer, say
$L$, exponentials, we can instead use the $L$ DFT coefficients with maximum absolute value.
Assume that $\psi(0),\ldots,\psi({L-1})$ are those coefficients.
Then our approximation $\tilw^{DFT}_L$ of $\omega$ by $L$ exponentials is given by:
\begin{equation}
\label{eqn_approx} \tilw^{DFT}_L(i)=\frac{1}{N}\sum_{k=0}^{L-1}
\psi(k) e^{\frac{2\pi \uimag}{N} k i} \quad \quad i = 0,\ldots,N-1.
\end{equation}

\eat{
The sequence of $N$ complex numbers $x_0,\ldots,x_{N-1}$ is transformed
into the sequence of $N$ complex numbers $X_0,\ldots,X_{N-1}$ as follows:

$$
X_k = \sum_{i=0}^{N-1} x_i e^{-\frac{2 \pi \uimag}{N} k i} \quad
\quad k = 0, \ldots, N-1
$$
where $\uimag$ is the imaginary unit and $e^{\frac{2 \pi \uimag}{N}}$  is a primitive $N$'th root of
unity.

The ''inverse discrete Fourier transform (IDFT)'' is given by
$$
x_i = \frac{1}{N} \sum_{k=0}^{N-1} X_k e^{\frac{2\pi \uimag}{N} k i}
\quad \quad i = 0,\ldots,N-1.
$$

We can see a discrete function $\omega(i)$
defined on a finite domain $[0,N-1]$ can be decomposed into $N$
exponentials exactly. If we want to approximate $\omega$ by fewer, say
$L$, exponentials, we can first do DFT on the sequence
$\omega(0),\ldots,\omega(N-1)$ and let $\psi(0),\ldots,\psi(N-1)$ be the resulting
sequence. For clarity, assume $\psi(0),\ldots,\psi(L-1)$ are the first $L$
DFT coefficients with maximum absolute values. Then our approximation
$\tilw^{DFT}_L$ of $\omega$ by $L$ exponentials will be given by
\begin{equation}
\label{eqn_approx} \tilw^{DFT}_L(i)=\frac{1}{N}\sum_{k=1}^{L}
\psi(k) e^{\frac{2\pi \uimag}{N} k i} \quad \quad i = 0,\ldots,N-1.
\end{equation}
}
\begin{figure}[t]
\center
\includegraphics[width=3in]{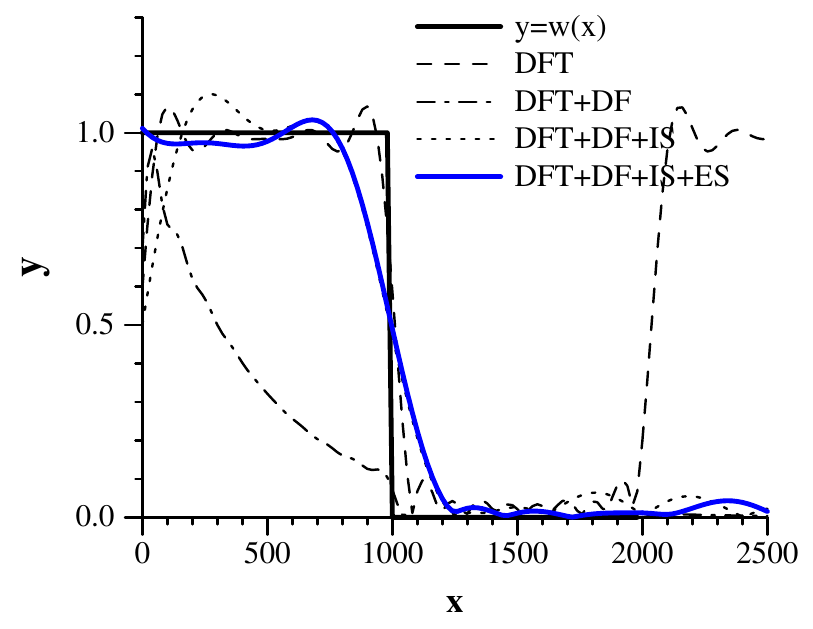}
\vspace{-5pt}
\caption{Illustrating the effect of the approximation steps: w(i) = step function with $N = 1000$, L = 20}
\vspace{-5pt}
\label{fig:approx_1}
\end{figure}

However, DFT utilizes only complex exponentials of unit norm, i.e., $e^{\uimag r}$ (where $r$ is a real), which makes this
approximation periodic (with a period of $N$).
This is not suitable for approximating an $\omega$ function used in PRF, which is typically a monotonically non-increasing function.
If we make $N$ sufficiently large, say larger than
the total number of tuples, then we usually need a large number of exponentials ($L$) to get a
reasonable approximation. Moreover, computing DFT for very large $N$ is computationally non-trivial.
Furthermore, the number of tuples $n$ may not be known in advance.

We next present a set of nontrivial tricks to adapt the base DFT approximation to overcome
these shortcomings. We assume $\omega(i)$ takes non-zero values within
interval $[0,N-1]$ and the absolute values of both $\omega(i)$ and
$\omega^{DFT}_L(i)$ are bounded by $B$.
To illustrate our method, we use the step function:
\[ \omega(i)=\left\{
                                                      \begin{array}{ll}
                                                        1, & \hbox{$i<N$} \\
                                                        0, & \hbox{$i\geq N$}
                                                      \end{array}
                                                    \right. \]
with $N = 1000$ as our running example to show our method and the specific shortcomings it addresses.
Figure \ref{fig:approx_1} illustrates the effect of each of these adaptations.

\begin{figure*}[t]
\center
\vspace{-5pt}
\includegraphics[width=.97\textwidth]{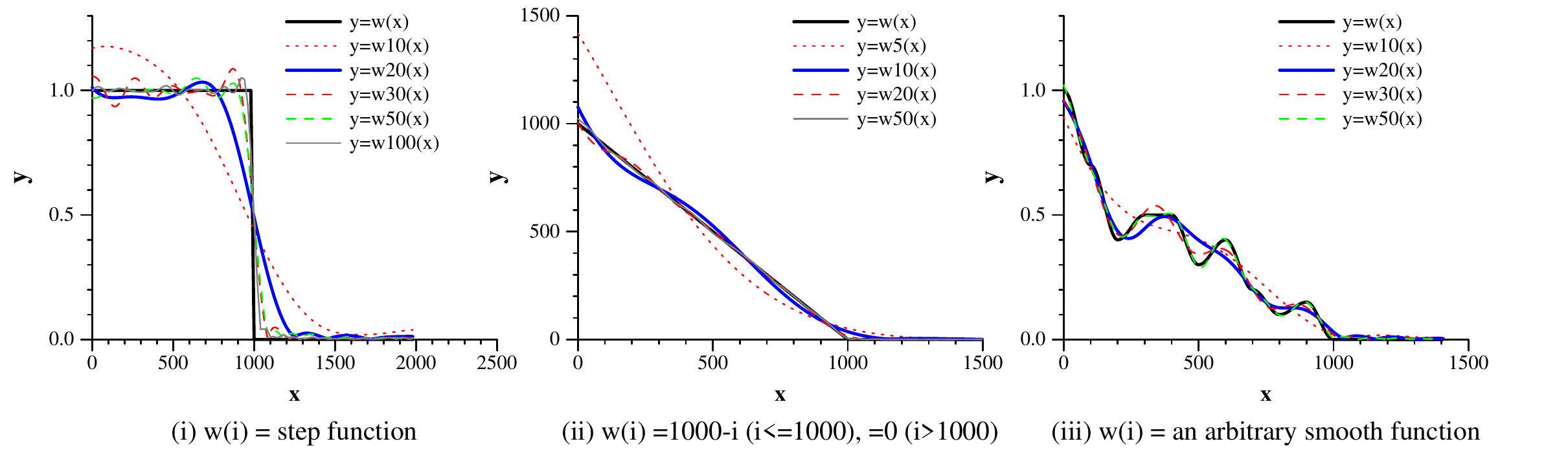}
\vspace{-5pt}
\caption{Approximating functions using linear combinations of complex exponentials: effect of increasing the number of coefficients}
\vspace{-10pt}
\label{fig:approx_2}
\end{figure*}

\begin{enumerate}
    \item {\bf (DFT)} We perform pure DFT on the domain $[1,a N]$, where $a$ is a small integer constant (typically $<10$).
        As we can see in Figure \ref{fig:approx_1} (where $N=1000$ and $a=2$), this results in a periodic approximation with a period of 2000.
        Although the approximation is reasonable for $x < 2000$, the periodicity is unacceptable if the number of tuples
        is larger than 2000 (since the positions between 2000 and 3000 (similarly, between 4000 and 5000) would be given high weights).

\smallskip
\smallskip
\item {\bf (Damping Factor (DF))} To address this issue, we introduce a damping factor $\eta\leq 1$ such that
$B\eta^{a N}\leq \epsilon$ where $\epsilon$ is a small positive
real (for example, $10^{-5}$). Our new approximation becomes:
\begin{equation}
\label{eqn_approx_df} \tilw^{DFT+DF}_L(i)=\eta^i\cdot \tilw^{DFT}_L(i)
=\frac{1}{N} \sum_{k=0}^{L-1} \psi(k) (\eta
e^{\frac{2\pi \uimag}{N} k})^i.
\end{equation}
By incorporating this damping factor, the periodicity is mitigated, since we have:
$\lim_{i\rightarrow +\infty}\tilw^{DFT+DF}_L(i)=0$. Especially,
$\tilw^{DFT+DF}_L(i)\leq \epsilon$ for $i> \alpha N$.

\smallskip
\smallskip
\item {\bf (Initial Scaling (IS))}
However the use of damping factor introduces another problem: it gives a biased approximation when $i$ is small (see Figure \ref{fig:approx_1}).
Taking the step function as an example, $\tilw^{DFT+DF}_L(i)$ is
approximately $\eta^i$ for $0\leq i<N$ instead of $1$. To rectify
this, we initially perform DFT on a different sequence
$\hat{\omega}(i)=\eta^{-i}\omega(i)$ (rather than $\omega(i)$) on domain $\in
[0,a N]$. Therefore, $\tilw^{DFT+IS}$ is a reasonable approximation of $\hat{\omega}$.
Then, if we apply the damping factor, it will give us an unbiased approximation of $\omega$,
which we denote by $\tilw^{DFT+DF+IS}$.

\smallskip
\smallskip
\item {\bf (Extending and Shifting (ES))}
This step is in particular tailored for optimizing the approximation performance for ranking functions.
DFT does not perform well at discontinuous points, specifically at $i=0$ (the left boundary), which can
significantly affect the ranking approximation. To handle this, we extrapolate $\omega$ to
make it continuous around $0$.
Let the resulting function be $\barw$ which is defined on $[-b N,
+\infty]$ for small $b>0$. Again, taking the step function for
example, we let $\barw(i)=\left\{
                                          \begin{array}{ll}
                                            1, & \hbox{$-b N\leq i< N$;} \\
                                            0, & \hbox{$i\geq N$.}
                                          \end{array}
                                        \right.$
Then, we shift $\barw(i)$ rightwards by $b N$ to make its
domain lie entirely in positive axis, do initial scaling and perform
DFT on the resulting sequence. We denote the approximation of the
resulting sequence by $\tilw'(i)$(by performing (\ref{eqn_approx_df})).
For the approximation of original $\omega(i)$ values, we only need to do
corresponding leftward shifting , namely
$\tilw^{DFT+DF+IS+ES}(i)=\tilw'(i+b N)$.
Figure \ref{fig:approx_1} shows that DFT+DF+IS+ES
gives a much better approximation than others around $i=0$.
\end{enumerate}

\noindent{Figures} \ref{fig:approx_1} and \ref{fig:approx_2}(i) illustrate the efficacy of our approximation technique for the step function.
As we can see, we are able to approximate that function
very well with just 20 or 30 coefficients.
Figure \ref{fig:approx_2}(ii) and (iii)
show the approximations for a piecewise linear function and
an arbitrarily generated continuous function respectively, both of which are
much easier to approximate than the step function.

\subsection{Learning a \PRFs\ or \PRFe\ Function}
\label{subsec_learnprfe}
\label{sec:learn prfe}
Next we address the question of how to learn the weights of a \PRFs\ function or the $\alpha$ for a single \PRFe\ function from user preferences.
To learn a linear combination of \PRFe\ functions, we first learn a \PRFs\ function and then approximate it as above.

Prior work on learning ranking functions (e.g.,~\cite{Herbrich98,Joachims02,RankNet,Dekel04}) assumes that the
user preferences are provided in the form of a set of pairs of tuples, and for each pair, we are told which
tuple is ranked higher. Our problem differs slightly from this prior work in that,
the features that we use to rank the tuples (i.e., $\Prob(r(t) = i), i=1, \dots, n$) cannot be computed for each tuple
individually, but must be computed for the entire dataset (since the values of the features for a tuple depend on the other
tuples in the dataset). Hence, we assume that we are instead given a small sample of the tuples, and
the user ranking for all those tuples. We compute the features assuming this sample
constitutes the entire relation, and learn a ranking function accordingly, with the goal to find the
parameters (the weights $w_i$ for \PRFs\ or the parameter $\alpha$ for \PRFe) that minimize
the number of disagreements with the provided ranking over the samples.

Given this, the problem of learning \PRFs\ is identical to the problem addressed in the prior work,
and we utilize the algorithm based on {\em support vector machines (SVM)}~\cite{Joachims02} in our experiments.

On the other hand, we are not aware of any work that has addressed learning a ranking
function like \PRFe. We use a simple binary search-like heuristic to find the optimal real value of $\alpha$
that minimizes the Kendall distance between the user-specified ranking and the ranking according to \PRFe($\alpha$).
In other words, we try to find $\arg\min_{\alpha\in [0,1]}(\dist(\sigma, \sigma(\alpha)))$ where
$\dist()$ is the Kendall distance between two rankings, $\sigma$ is the ranking for the given sample
and $\sigma(\alpha)$ is the one obtained by using \PRFe($\alpha$) function.
Suppose we want to find the optimal $a$ within the interval $[L,U]$
now. We first compute $\dist(\sigma, \sigma(L+i\cdot{U-L\over 10})$
for $i=1,\ldots,9$ and find $i$ for which the distance is the
smallest. Then we reduce our search range to
$[\max(L,L+(i-1)\cdot{U-L\over 10},\min(U,L+(i+1)\cdot{U-L\over
10})]$ and repeat the above recursively.
Although this algorithm can only converge to a local minimum, in our experimental study, we observed
that all of the prior ranking functions exhibit a uni-valley behavior (Section \ref{sec:experiments}),
and in such cases, this algorithm finds
the global optimal.

\section{\PRF\ as a Consensus Top-$k$ Answer}
\label{sec:contopk}
\label{sec:contop-k}
In this section, we show there is a close connection between \PRFs\ and
the notion of consensus top-$k$ answer (\CON) proposed in \cite{pods09_LD}.
We first review the definition of a consensus top-$k$ ranking.
\begin{definition}
Let $\dist()$ denote a distance function between two top-$k$ rankings.
Then the {\em most consensus answer} $\tau$
is defined to be the top-$k$ ranking such that the expected distance between $\tau$
and the answer $\tau_{pw}$ of the (random) world $pw$ is minimized,
i.e.,
$$
\tau=\arg\min_{\tau'\in \Omega}\{\Exp[\dist(\tau',\tau_{pw})]\}.
$$
\end{definition}
$\dist()$ may be any distance function defined on pairs of top-$k$ answers.
In \cite{pods09_LD}, we discussed how to compute or approximate \CON\ under a number of distance functions,
such as Spearman's rho, Kendall's tau and intersection metric~\cite{fagin:soda03}.

\begin{example}
Consider the example in Figure~\ref{fig_example}.
Assume $k=2$ and the distance function is the symmetric difference metric
$\dist_{\Delta}=|(\tau_1\backslash \tau_2)\cup (\tau_2\backslash \tau_1)|$.
The most consensus top-$2$ answer is $\{t_{2}, t_{5}\}$ and the expected distance is
$\Exp[\dist(\tau',\tau_{pw})]=.112\times 2 + .168\times 2 +.048\times 4+.072\times 4 +.168\times 2+.252\times 0 +.072\times 4+.108\times 2$.
\end{example}

We first show that a \CON\ answer under symmetric difference is equivalent to \PT($k$), a special case
of \PRFs. Then, we generalize the result and show that any \PRFs\ function is in fact equivalent to
some \CON\ answer under some suitably defined distance function that generalizes
symmetric difference.
This new connection further justifies the semantics of \PRFs\ from an optimization point of view
in that the top-$k$ answer obtained by \PRFs\ minimizes the expected value of some distance function,
and it may shed some light on designing the weight function for \PRFs\ in particular applications.

\subsection{Symmetric Difference and PT-k Ranking Function}

Recall \PT($k$) query returns
$k$ tuples with the largest $\Prob(r(t)\leq k)$ values.
We show that the answer returned is the \CON\ under symmetric difference metric $\dist_{\Delta}$
where $
\dist_{\Delta}(\tau_1,\tau_2)=|\tau_1\Delta
\tau_2|=|(\tau_1\backslash \tau_2)\cup (\tau_2\backslash \tau_1)|
$
\footnote{The result of this subsection has appeared in \cite{pods09_LD}.}.

For ease of notation, we let $\Prob(r(t)>i)$ includes the probability that $t$'s rank
is larger than $i$ and that $t$ doesn't exist.
We use the symbol $\tau$ to denote
a top-k ranked list. We use $\tau(i)$ to denote the $i^{\text{th}}$ item in the list $\tau$ for positive integer $i$,
and $\tau(t)$ to denote the position of $t\in T$ in $\tau$.

\begin{theorem}
\label{thm_mindis_prfk}
If $\tau=\{\tau(1),\tau(2),\ldots,\tau(\rmk)\}$ is the set of
$\rmk$ tuples with the largest $\Prob(r(t)\leq \rmk)$,
then $\tau$ is the \CON\ answer under metric $\dist_{\Delta}$, i.e., the answer minimizes $\Exp[\dist_{\Delta}(\tau,\tau_{pw})]$.
\end{theorem}
\begin{proof}
Suppose $\tau$ is fixed.
We write $\Exp[\dist_{\Delta}(\tau,\tau_{pw})]$ as follows:
\begin{align*}
\Exp\,[\dist_{\Delta}(\tau,\tau_{pw})]&= \Exp\,\Bigl[\sum_{t\in T} \delta(t\in \tau \wedge t\notin \tau_{pw})+\delta(t\in \tau_{pw} \wedge t\notin \tau)\Bigr]  \\
&= \sum_{t\in T\setminus \tau} \Exp[\delta(t\in \tau_{pw})] +\sum_{t\in \tau}  \Exp[\delta(t\notin \tau_{pw})] \\
&= \sum_{t\in T\setminus\tau} \Prob(r(t)\leq \rmk) +\sum_{t\in \tau} \Prob(r(t)> \rmk) \\
&= \rmk + \sum_{t\in T} \Prob(r(t)\leq\rmk) - 2\sum_{t\in \tau} \Prob(r(t)\leq \rmk)
\end{align*}
The first two terms are invariant with respect to $\tau$.
Therefore, it is clear that the set of $\rmk$ tuples with the largest $\Prob(r(t)\leq \rmk)$
minimizes the expectation.
\qed
\end{proof}

\subsection{Weighted Symmetric Difference and \PRFs}

We present a generalization of Theorem~\ref{thm_mindis_prfk} that shows
the equivalence between any \PRFs\ function
and \CON\ under {\em weighted symmetric difference} distance functions which generalize the symmetric difference.
Suppose $\omega$ is a positive function defined on $\mathbb{Z}^{+}$ and $\omega(i)=0 \forall i>k$.
\begin{definition}
The weighted symmetric difference with weight $\omega$ of two top-$k$ answers $\tau_{1}$ and $\tau_{2}$
is defined to be
$$\dist_{\omega}(\tau_{1},\tau_{2})=\sum_{i=1}^{k} \omega(i)\delta(\tau_{2}(i)\notin \tau_{1}).$$
\end{definition}
Intuitively, if the $i^{\text{th}}$ item of $\tau_{2}$ can not be found in $\tau_{1}$, we pay a penalty of $\omega(i)$ and
the distance is just the total penalty.
If $\omega$ is a decreasing function,  the distance function captures the intuition that top ranked items should carry more weight.
If $\omega$ is a constant function, it reduces to the ordinary symmetric difference distance.
Note that $\dist_{\omega}$ is not necessarily symmetric
\footnote{Rigorously, a distance function (or metric) should satisfy positive definiteness,
symmetry and triangle inequality. Here we abuse this term a bit.}.
Now, we present the theorem which is a generalization of Theorem~\ref{thm_mindis_prfk}.

\begin{theorem}
\label{thm_mindis_prfk2}
Suppose $\omega$ is a positive function defined on $\mathbb{Z}^{+}$ and $\omega(i)=0 \forall i>k$.
If $\tau=\{\tau(1),\tau(2),\ldots,\tau(\rmk)\}$ is the set of
$\rmk$ tuples with the largest $\rank_{\omega}(t)$ values,
then $\tau$ is the \CON\ answer under the weighted symmetric difference $\dist_{\omega}$, i.e., the answer minimizes $\Exp[\dist_{\omega}(\tau,\tau_{pw})]$.
\end{theorem}
\begin{proof}
The proof mimics the one for Theorem~\ref{thm_mindis_prfk}.
Suppose $\tau$ is fixed.
We can write $\Exp[\dist_{\omega}(\tau,\tau_{pw})]$ as follows:
\begin{align*}
\Exp\,[\dist_{\omega}(\tau,\tau_{pw})]&= \Exp\,\Bigl[\sum_{t\in T} \omega(\tau_{pw}(t)) \delta(t\in \tau_{pw} \wedge t\notin \tau)\Bigr]  \\
&= \sum_{t\in T\setminus \tau} \Exp[\omega(\tau_{pw}(t))\delta(t\in \tau_{pw})] \\
&= \sum_{t\in T\setminus\tau} \sum_{i=1}^{k} \omega(i) \Prob(r(t)=i) =\sum_{t\in T\setminus\tau} \rank_{\omega}(t)
\end{align*}
Therefore, it is clear that the set of $\rmk$ tuples with the largest $\rank_{\omega}(t)$ values
minimizes the above quantity.
\qed
\end{proof}

Although the weighted symmetric difference appears to be a very rich class of distance functions,
its relationship with other well studied distance functions, such at Spearman's rho and Kendall's tau,
is still not well understood. We leave it as an interesting open problem.

\section{An Interesting Property of \PRFe}
\label{sec:prfeprop}
We have seen that \PRFe($\alpha$) admits very efficient evaluation algorithms.
We also suggest that the parameter $\alpha$ should be learned from samples or user feedback.
In fact, we do so since since we hold the promise that
by changing the parameter $\alpha$, \PRFe\ can span a spectrum of rankings, and the true ranking should be part
of this spectrum or close to some point in it.
We provide empirical support for this claim shortly in the next section (Section~\ref{sec:experiments}).
In this section, we make some interesting theoretical observations about \PRFe, which help us further
understand the behavior of \PRFe\ itself.

First, we observe that for $\alpha = 1$, the \PRFe\ ranking is equivalent to the ranking of tuples by their existence probabilities
(\PRFe\ value in that case is simply the total probability).
On the other hand, when $\alpha$ approaches $0$, \PRFe\ tends to rank
the tuples by their probabilities to be the top-$1$ answer, i.e., $\Prob(r(t)=1)$.
Thus, it is a natural question to ask how the ranking changes when we vary $\alpha$ from $0$ to $1$.
Now, we prove the following theorem which gives an important characterization of the behavior of \PRFe\
on tuple independent databases.

Let $\tau_{\alpha}$ denote the ranking obtained by \PRFe($\alpha$). For simplicity, we ignore the possibility of ties and
assume this ranking is unique.
As two special cases,
let $\tau_{0}$ and $\tau_{1}$ denote the rankings obtained by sorting the tuples in a decreasing $\Prob(r(t)=1)$ and $\Prob(t)$ order,
respectively.

\begin{theorem}
\begin{enumerate}
\item If $t_{i} >_{\tau_{0}} t_{j}$ ($t_{i}$ is ranked higher than $t_{j}$ in $\tau_{0}$)
and  $t_{i} >_{\tau_{1}} t_{j}$, then $t_{i}>_{\tau_{\alpha}}t_{j}$  any $0\leq \alpha\leq 1$.
\item If $t_{i} >_{\tau_{0}} t_{j}$
and  $t_{i} <_{\tau_{1}} t_{j}$, then there is exactly one point $\beta$ such that
$t_{i}>_{\tau_{\alpha}}t_{j}$ for $\alpha<\beta$ and
$t_{i}<_{\tau_{\alpha}}t_{j}$ for $\alpha>\beta$.
\end{enumerate}
\end{theorem}
\begin{proof}
Let $\rank_{\alpha}(t_{i})$ be the \PRFe($\alpha$) value of tuple $t_{i}$.
Then:
\begin{align*}
\rank_{\alpha}(t_i)=\calF^i(\alpha)=\biggl(\prod_{t\in T_{i-1}} \bigl(1-\Prob(t)+\Prob(t)\alpha \bigr)\biggr)\Prob(t_i)\alpha.
\end{align*}
Assume that $i<j$. Dividing $\rank_{\alpha}(t_{j})$ by $\rank_{\alpha}(t_{i})$, we get
\begin{align*}
\rho_{j,i}(\alpha)&=\frac{\rank_{\alpha}(t_j)}{\rank_{\alpha}(t_i)}=\frac{\prod_{t\in T_{j-1}} \bigl(1-\Prob(t)+\Prob(t)\alpha \bigr)}
{\prod_{t\in T_{i-1}} \bigl(1-\Prob(t)+\Prob(t)\alpha \bigr)} \cdot \frac{\Prob(t_{j})}{\Prob(t_{i})}\\
&=\frac{\Prob(t_{j})}{\Prob(t_{i})} \cdot
\prod_{l=i}^{j-1} \bigl(1-\Prob(t_{l})+\Prob(t_{l})\alpha \bigr)
 \end{align*}
Notice that $1-\Prob(t)+\Prob(t)\alpha$ is always non-negative and an increasing function of $\alpha$.
Therefore, $\rho_{j,i}(\alpha)$ is increasing in $\alpha$.
If $i>j$, the same argument show $\rho_{j,i}(\alpha)$ is decreasing in $\alpha$.
In either case,  the ratio is monotone in $\alpha$.

If $\rho_{j,i}(0)<1$ and $\rho_{j,i}(1)<1$, then $\rho_{j,i}(\alpha)<1$ for all $0<\alpha\leq 1$.
Therefore, the first half of the theorem holds.
If $\rho_{j,i}(0)<1$ and $\rho_{j,i}(1)>1$, then there is exactly one point $0<\beta<1$ such that $\rho_{j,i}(\beta)=1$,
$\rho_{j,i}(\alpha)<1$ for all $0<\alpha<\beta$,
and $\rho_{j,i}(\alpha)>1$ for all $\beta<\alpha\leq 1$.
This proves the second half. \qed
\end{proof}

Some nontrivial questions can be immediately answered by the theorem.
For example, one may ask the question ``Is it possible that we get some ranking $\tau_{1}$,
increase $\alpha$ a bit and get another ranking $\tau_{2}$, and increase $\alpha$ further and get $\tau_{1}$ back?'',
and we can quickly see that the answer is no; if two tuples change positions, they never change back.
Another observation we can make is that: if
$t_{1}$ dominates $t_{2}$ (i.e., $t_{1}$ has a higher score and probability), then $t_{1}$ always ranks above
$t_{2}$ for any $\alpha$ (this is because $t_{1}$ ranks above $t_{2}$ in both $\tau_{0}$ and $\tau_{1}$).

Interestingly, the way the ranking changes as $\alpha$ is increased from 0 to 1 is reminiscent of the execution of the {\em bubble sort algorithm}.
We assume the true order of the tuples is $\tau_{1}$ and the initial order is $\tau_{0}$.
We increase $\alpha$ from $0$ to $1$ gradually. Each change in the ranking is just a swap of a pair of adjacent tuples
that are not in the right relative order initially. The number of swaps is exactly the number of reversed pairs.
This is just like bubble sort! The only difference is that the order of those swaps may not be the same.

\begin{figure}[t]
    \vspace{-5pt}
\center{
\includegraphics[width=0.4\linewidth]{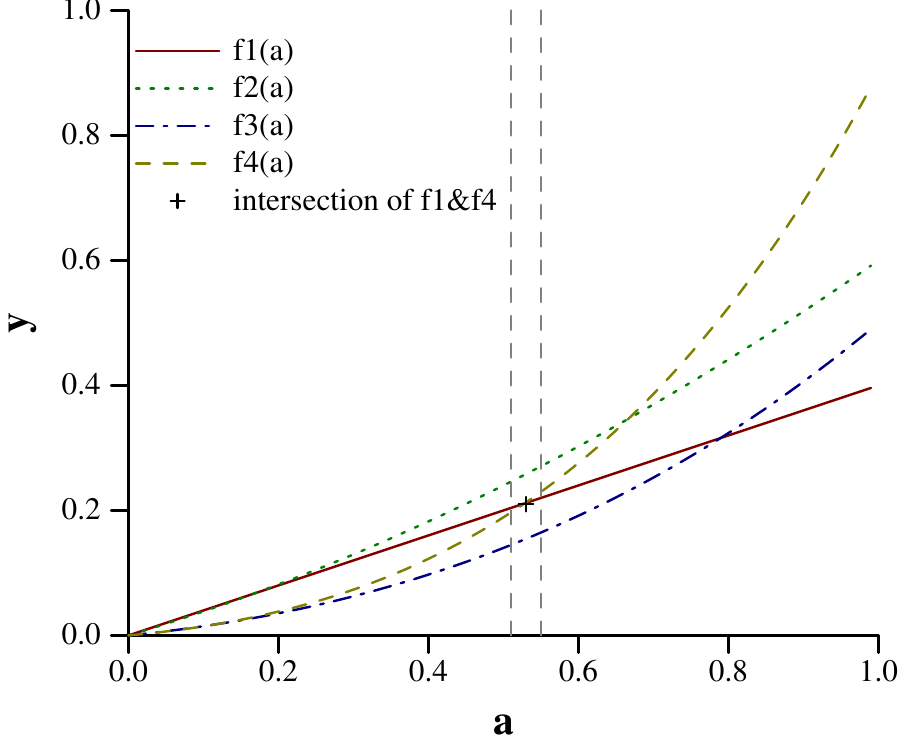}
}
\caption{Illustration of Example~\ref{ex:change}.
$f_{i}(\alpha)=\rank_{\alpha}(t_{i})$ for $i=1,2,3,4$.
}
\vspace{-10pt}
\label{eg_change}
\end{figure}

\begin{example}
\label{ex:change}
Suppose we have four independent tuples: \\
 \centerline{$(t_{1}: 100, .4),(t_{2}: 80, .6), (t_{3}: 50, .5), (t_{4}: 30, .9)$} \\
Using (\ref{eqn_prfe}), it is easy to see that
$\rank_{\alpha}(t_{1})=.4\alpha, \rank_{\alpha}(t_{2})=(.6+.4\alpha).6\alpha,
\rank_{\alpha}(t_{3})=(.6+.4\alpha)(.4+.6\alpha).5\alpha$
and $\rank_{\alpha}(t_{4})=(.6+.4\alpha)(.4+.6\alpha)(.5+.5\alpha).9\alpha$.
In Figure~\ref{eg_change}, each curve corresponds to one tuple.
In interval $(0,1]$, any two curves intersect at most once.
Changes in the ranking happen right at the intersection points and one adjacent pair
of tuples swap their positions. For instance, the $+$ sign in the figure is the intersection point
of $f_{1}$ and $f_{4}$. The rank list is $\{t_{2}, t_{1}, t_{4}, t_{3}\}$ right before the point
and  $\{t_{2}, t_{4}, t_{1}, t_{3}\}$ right after the point.
\end{example}

In fact, if we think of $h$ as a parameter of \PT\ and we vary $h$ from $1$ to $n$,
the process that the rank list changes is quite similar to the one for \PRFe:
On one extreme where $h=1$, the rank list is $\tau_{0}$, i.e., the tuples are sorted by $\Prob(r(t)=1)$ and
on the other extreme where $h=n$, the rank list is $\tau_{1}$, i.e., the tuples are sorted by $\Prob(r(t)\leq n)=\Prob(t)$.
 However, \PT\ is only able to explore at most $n$ different rankings (one for each $h$)
 ``between'' $\tau_{0}$ and $\tau_{1}$,
 while \PRFe\ may explore $O(n^{2})$ of them.

\section{Experimental Study}
\label{sec_experiment}
\label{sec:experiments}

We conducted an extensive empirical study over several real and synthetic datasets to
illustrate: (a) the diverse and conflicting behavior of different ranking functions proposed in the
prior literature, (b) the effectiveness of our parameterized ranking functions, especially \PRFe, at
approximating other ranking functions, and (c) the scalability of our new generating functions-based
algorithms for exact and approximate ranking. We discussed the results supporting (a) in
Section~\ref{sec:prior work}.
In this section, we focus on (b) and (c).

\topic{Datasets} We mainly use the International Ice Patrol (IIP) Iceberg Sighting Dataset\footnote{ http://nsidc.org/data/g00807.html} for our experiments.
This dataset was also used in prior work on ranking in probabilistic databases~\cite{Jin08_sliding,conf/sigmod/HuaPZL08}.
The database contains a set of {\em iceberg sighting records}, each of which contains
the location ({\em latitude, longitude}) of the iceberg, and the {\em number of days} the iceberg has drifted, among other attributes.
Detecting the icebergs that have been drifting for long periods is crucial, and hence
we use the number of days drifted as the ranking score.
The sighting record is also associated with a {\em confidence-level} attribute
according to the source of sighting: R/V (radar and visual), VIS (visual only), RAD (radar only),
SAT-LOW (low earth orbit satellite), SAT-MED (medium earth orbit
satellite), SAT-HIGH (high earth orbit satellite), and EST (estimated).
We converted these six confidence levels into probabilities 0.8, 0.7, 0.6, 0.5, 0.4, 0.3, and 0.4 respectively.
We added a very small Gaussian noise to each probability so that ties could be broken.
There are nearly a million records available from 1960 to 2007; we created 10 different datasets for our experimental
study containing $100,000$ (IIP-100,000) to $1,000,000$ (IIP-1,000,000) records, by uniformly sampling from
the original dataset.

Along with the real datasets, we also use several synthetic datasets with varying
degrees of correlations, where the correlations are captured using probabilistic and/xor trees.
The tuple scores (for ranking) were chosen uniformly at random from $[0,10000]$. The
corresponding and/xor trees were also generated randomly starting with the root, by controlling
the {\em height (L)}, the {\em maximum degree of the non-root nodes (d)}, and the {\em proportion of $\Cvee$
and $\Cwedge$ nodes (X/A)} in the tree. Specifically, we use five such datasets:
\begin{enumerate}
\item[1.] Syn-IND (independent tuples): the tuple existence probabilities were chosen uniformly at random from $[0,1]$.
\item[2.] Syn-XOR (L=2,X/A=$\infty$,d=5): Note that the Syn-XOR dataset, with height set to 2 and no $\Cwedge$ nodes,
exhibits only mutual exclusivity correlations (mimicking the x-tuples model~\cite{sarma:icde06,conf/icde/YiLSK08})
\item[3.] Syn-LOW (L=3,X/A=10,d=2)
\item[4.] Syn-MED (L=5,X/A=3,d=5)
\item[5.] Syn-HIGH (L=5,X/A=1,d=10).
\end{enumerate}

\topic{Setup} We use the normalized Kendall distance (Section \ref{sec:distance-topk}) for comparing two top-k rankings. All the algorithms
were implemented in C++, and the experiments were run on a 2.4GHz Linux PC with 2GB memory.

\begin{figure*}[t]
\begin{center}
\includegraphics[width=0.9\linewidth]{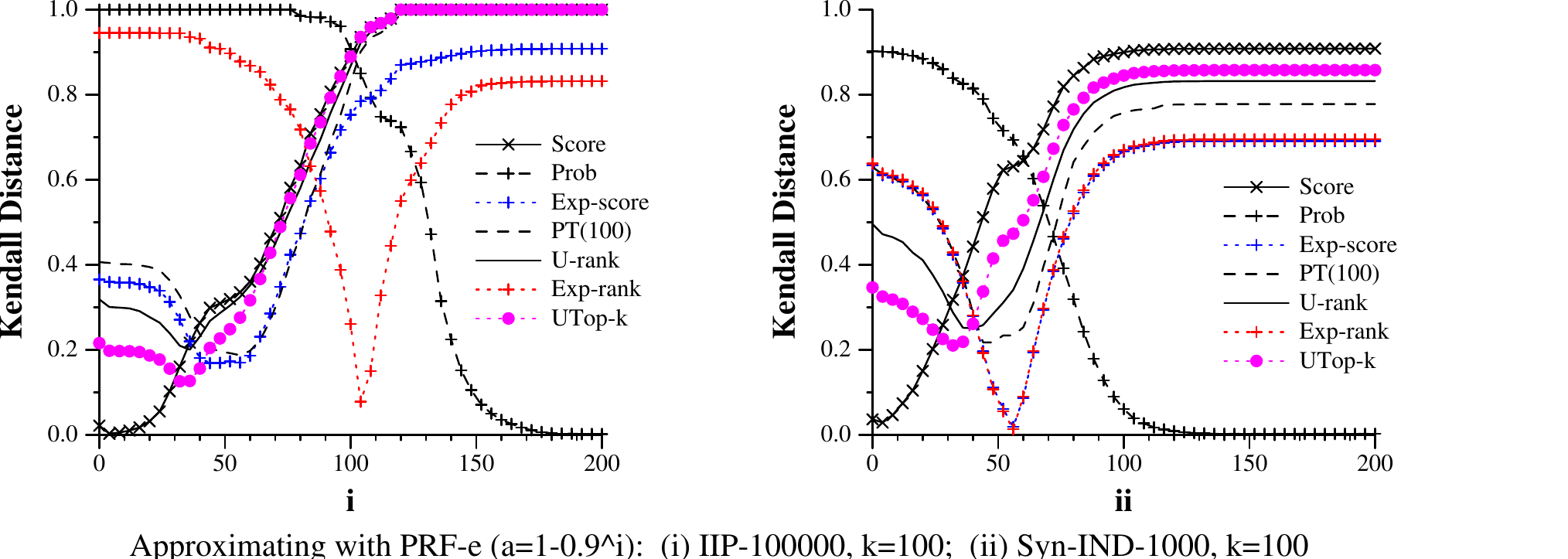}
\end{center}
\caption{Comparing \PRFe\ with other ranking functions for varying values of $\alpha$;
        (i))IIP-100,000, (ii)Syn-IND-1000}
\label{fig:approximating_1}
\end{figure*}

\subsection{Approximability of Ranking Functions}
\label{subsec_exp_approx}
We begin with a set of experiments illustrating the effectiveness of our parameterized ranking functions at approximating
other ranking functions. Due to space constraints, we focus on \PRFe\ here because it is significantly faster to rank according
to a \PRFe\ function (or a linear combination of several \PRFe\ functions) than it is to rank according a \PRFs\ function.

Figures \ref{fig:approximating_1} (i) and (ii) show the Kendall distance between the Top-100 answers computed using
a specific ranking function and \PRFe\ for varying values of $\alpha$, for the IIP-100,000 and Syn-IND-1000 datasets. For
better visualization, we plot $i$ on the x-axis, where $\alpha = 1 - 0.9^i$. The reason behind this is that
the behavior of the \PRFe\ function changes rather drastically, and spans a spectrum of rankings,
when $\alpha$ approaches $1$. First, as we can see, the \PRFe\ ranking is close to ranking by {\em Score} alone
for small values of $\alpha$, whereas it is close to the ranking by {\em Probability} when $\alpha$ is close
to 1 (in fact, for $\alpha = 1$, the \PRFe\ ranking is equivalent to the ranking of tuples by their existence probabilities)\footnote{On the other hand, for $\alpha = 0$, \PRFe\ ranks
the tuples by their probabilities to be the Top-$1$ answer.}. 
Second, we see that, for all other functions (\ES, \PT, \URK, \ERK),
there exists a value of $\alpha$ for which the distance of that function to
\PRFe\ is very small, indicating that \PRFe\ can indeed approximate those functions quite well.
Moreover we observe that this ``uni-valley'' behavior of the curves justifies
the binary search algorithm we advocate for learning the value of $\alpha$
in Section~\ref{subsec_learnprfe}. Our experiments with other synthetic and real datasets indicated
a very similar behavior by the ranking functions.

\begin{figure*}[t]
\center{\includegraphics[width=0.9\linewidth]{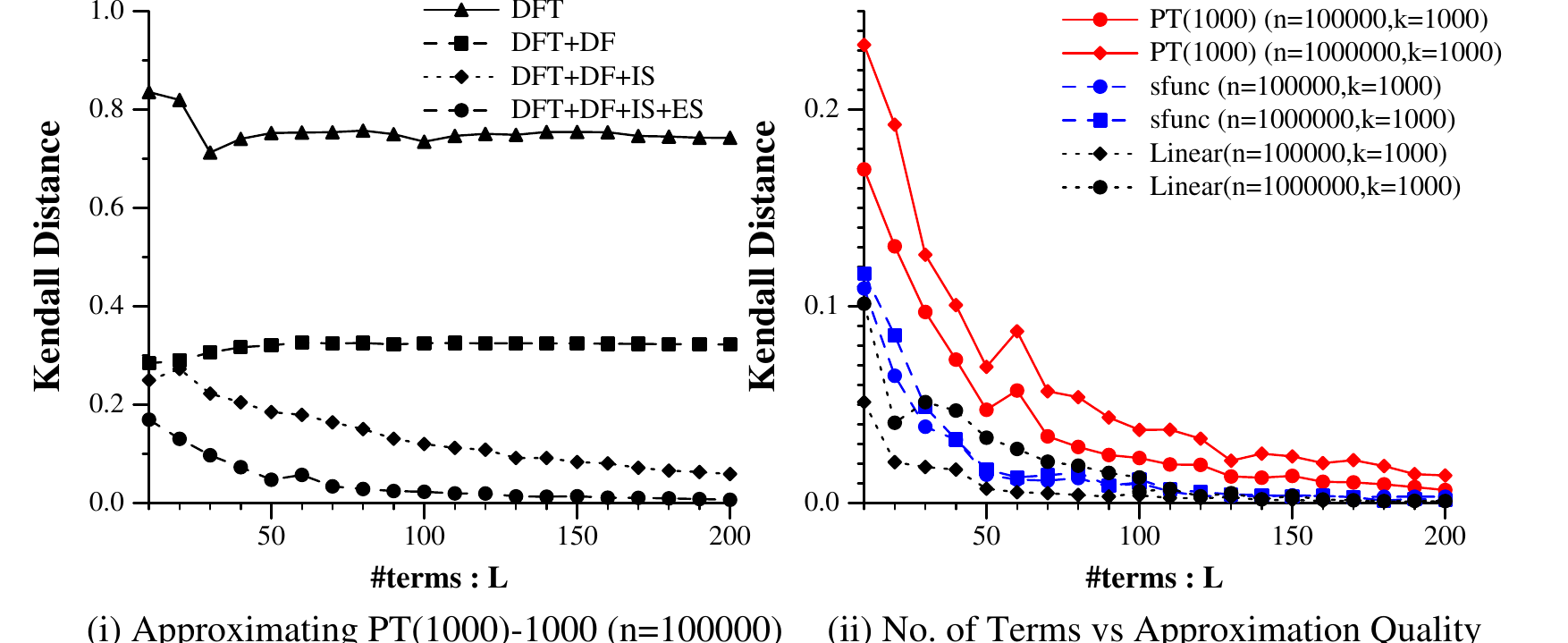}}
\caption{(i) Approximating \PTPARAM{1000}
    using a linear combination of \PRFe\ functions; (ii) Approximation quality for three ranking functions for varying number
        of exponentials.}
\label{fig:approximating_2}
\end{figure*}

Next we evaluate the effectiveness of our approximation technique presented 
in Section~\ref{sec:approximating and learning}.
In Figure \ref{fig:approximating_2} (i), we show the Kendall distance between the top-k answers
obtained using \PT\ (for $h=1000,\rmk=1000$) and using a linear combination of \PRFe\ functions found by our algorithms.
As expected, the approximation using the vanilla DFT technique is very bad, with the Kendall distance close to 0.8
indicating little similarity between the \Topk\ answers. However, the approximation obtained using our proposed
algorithm (indicated by DFT+DF+IS+ES curve) achieves a Kendall distance of less than 0.1 with just $L=20$ exponentials.

In Figure \ref{fig:approximating_2} (ii), we compare the approximation quality (found by our algorithm DFT+DF+IS+ES)
for three ranking functions for two datasets: IIP-100,000 with $\rmk=1000$, and IIP-1,000,000 dataset with $\rmk=10000$.
The ranking functions we compared were: (1) \PT\ ($h=1000$), (2) an arbitrary smooth function, $sfunc$, 
and (3) a linear function (Figure \ref{fig:approximating_2}(ii)).
We see that $L=40$ suffices to bring the Kendall distance to $< 0.1$ in all cases.
We also observe that smooth functions (for which the absolute value of the first derivative
of the underlying continuous function is bounded by a small value) are usually easier to
approximate. We only need $L=20$ exponentials to achieve a Kendall distance less than $0.05$
for $sfunc$. The Linear function is even easier to approximate.

\subsection{Learning Ranking Functions}
Next we consider the issue of learning ranking functions from user preferences.
Lacking real user preference data, we instead assume that the user ranking function, denoted {\em user-func},
is identical to one of: \ES, \PT, \URK, \ERK, or \PRFe($\alpha=0.95$).  We generate
a set of user preferences by ranking a random sample of the dataset using {\em user-func} (thus generating five
sets of user preferences). These are then fed to the learning algorithm, and finally we compare the Kendall distance
between the learned ranking and the true ranking for the entire dataset.

\begin{figure*}[t]
\begin{center}
    \includegraphics[width=0.9\textwidth]{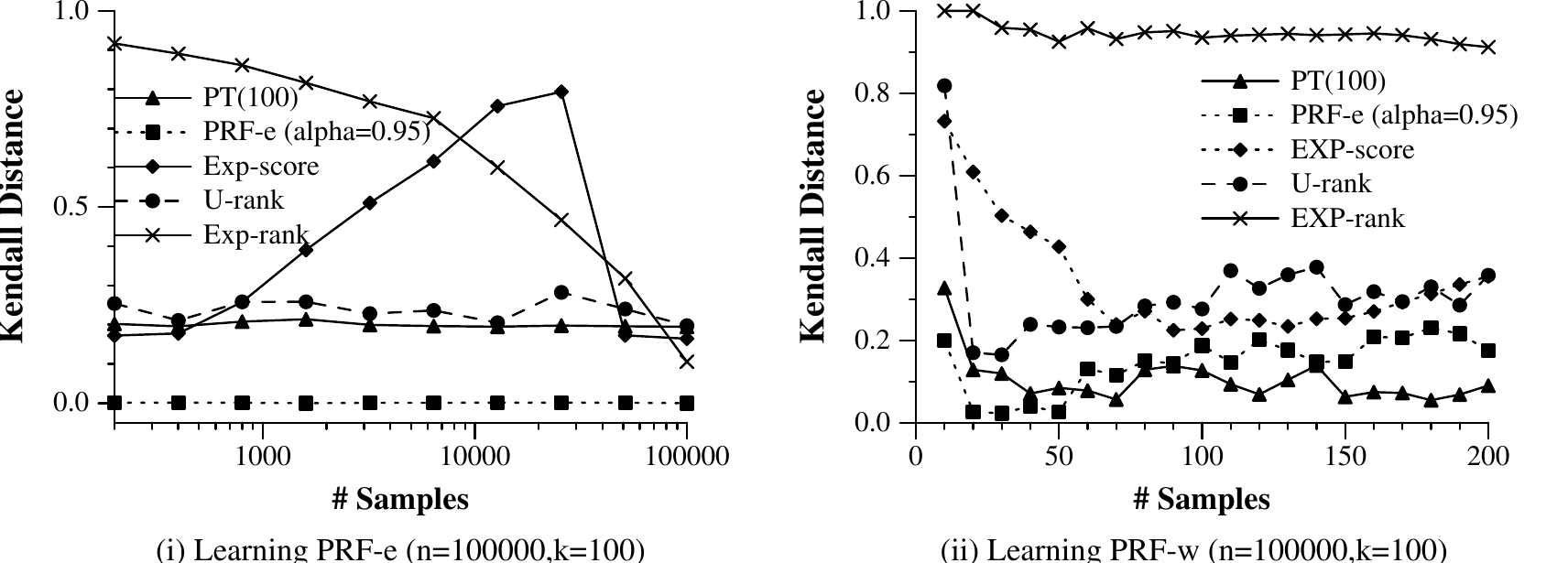}
\end{center}
\caption{(i) Learning \PRFe\ from user preferences; (ii) Learning \PRFs\ from user preferences.}
\label{fig:learning-correlations_1}
\end{figure*}

In Figure \ref{fig:learning-correlations_1}(i), we plot the results for learning a single \PRFe\ function
(i.e., for learning the value of $\alpha$) using the binary search-like algorithm presented in Section \ref{sec:learn prfe}.
The experiment reveals that when the underlying ranking is done by \PRFe, the value of $\alpha$ can be learned
perfectly. When one of \PT\ or \URK\ is the underlying ranking function,
the correct value $a$ can be learned with a fairly small sample size, and increasing the number
of samples does not help in finding a better $\alpha$.
On the other hand, \ERK\ cannot be learned well by \PRFe\ unless the sample size approaches
the total size of whole dataset. This phenomenon can be partly explained using Figure \ref{fig:approximating_1}(i) and (ii) in which
the curves for \PT\ and \UTK\ have a fairly smooth valley,
while the one for \ERK\ is very sharp and the region of $\alpha$ values
where the distance is low is extremely small ($[1-0.9^{90},1-0.9^{110}]$).
Hence, the minimum point for \ERK\ is harder to reach.
Another reason is that \ERK\ is quite sensitive to the size of the dataset, which
makes it hard to learn it using a smaller-sized sample dataset.
We also observe that while extremely large samples are able to learn \ES\ well, the behavior of \ES\ is quite
unstable when the sample size is smaller.

Note that if we already know the form of the ranking function, we don't need to learn it in this fashion; we can
instead directly find an approximation for it using our DFT-based algorithm.

In Figure \ref{fig:learning-correlations_1} (ii), we show the results of an experiment
where we tried to learn a \PRFs\ function (using the SVM-lite package~\cite{Joachims02}).
We keep our sample size $\leq 200$ since
SVM-lite becomes drastically slow with larger sample sizes. 
First we observe that \PT\ and \PRFe\ can be learned very well from a small size sample (distance $<0.2$ in most cases)
and increasing the sample size does not benefit significantly. \URK\ can also be learned, but the approximation isn't nearly as good.
This is because \URK\ can not be written as a single \PRFs\ function. We observed similar behavior in our experiments
with other datasets. Due to space constraints, we omit a further discussion on learning a \PRFs\ function; the issues in
learning such weighted functions have been investigated in prior literature, and if the true
ranking function can be written as a \PRFs\ function, then the above algorithm is expected to learn it well given
a reasonable number of samples.

\begin{figure*}[t]
\begin{center}
    \includegraphics[width=0.9\textwidth]{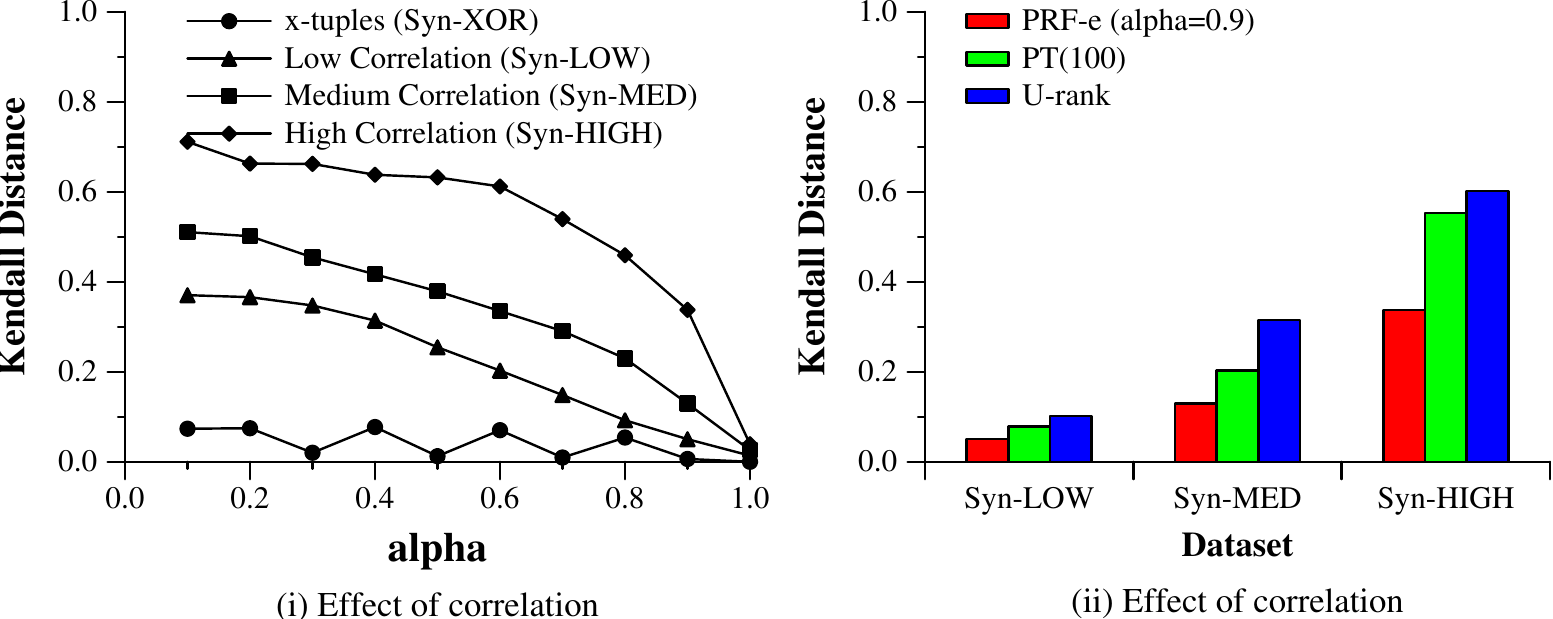}
\end{center}
\caption{(i) Effect of correlations on \PRFe\ ranking as $a$ varies;
(ii) Effect of correlations on \PRFe, \URK\ and \PT.}
\label{fig:learning-correlations_2}
\end{figure*}

\subsection{Effect of Correlations}
Next we evaluate the behavior of ranking functions over probabilistic datasets modeled using probabilistic and/xor trees.
We use the four synthetic correlated datasets, Syn-XOR, Syn-LOW, Syn-MED, and Syn-HIGH, for these experiments.
For each dataset and each ranking function considered, we compute the rankings by considering the correlations, and by ignoring the correlations,
and then compute the Kendall distance between these two (e.g., for \PRFe, we compute the rankings using \textbf{PROB-ANDOR-PRF-RANK} and
\textbf{IND-PRF-RANK} algorithms). Figure \ref{fig:learning-correlations_2}(i) shows the results for the \PRFe\ ranking function for
varying $\alpha$, whereas in Figure \ref{fig:learning-correlations_2}(ii), we plot the results for \PRFe($\alpha=0.9$), \PTPARAM{100}, and \URK.

As we can see, on highly correlated datasets, ignoring the correlations can result in significantly
inaccurate \Topk\ answers. This is not as pronounced for the Syn-XOR dataset. This is because, in any group of tuples that
are mutually exclusive, there are typically only a few tuples that may have sufficiently high probabilities
to be part of the top-k answer; the rest of the tuples may be ignored for ranking purposes. Because of this,
assuming tuples to be independent of each other does not result in significant errors.
As $\alpha$ approaches $1$, \PRFe\ tends to sort the tuples by probabilities,
so all four curves in Figure \ref{fig:learning-correlations_2}(i) become close to $0$. We note that
ranking by \ES\ is invariant to the correlations, which is a significant drawback of that function.


\begin{figure*}[t]
    \centerline{\includegraphics[width=0.9\textwidth]{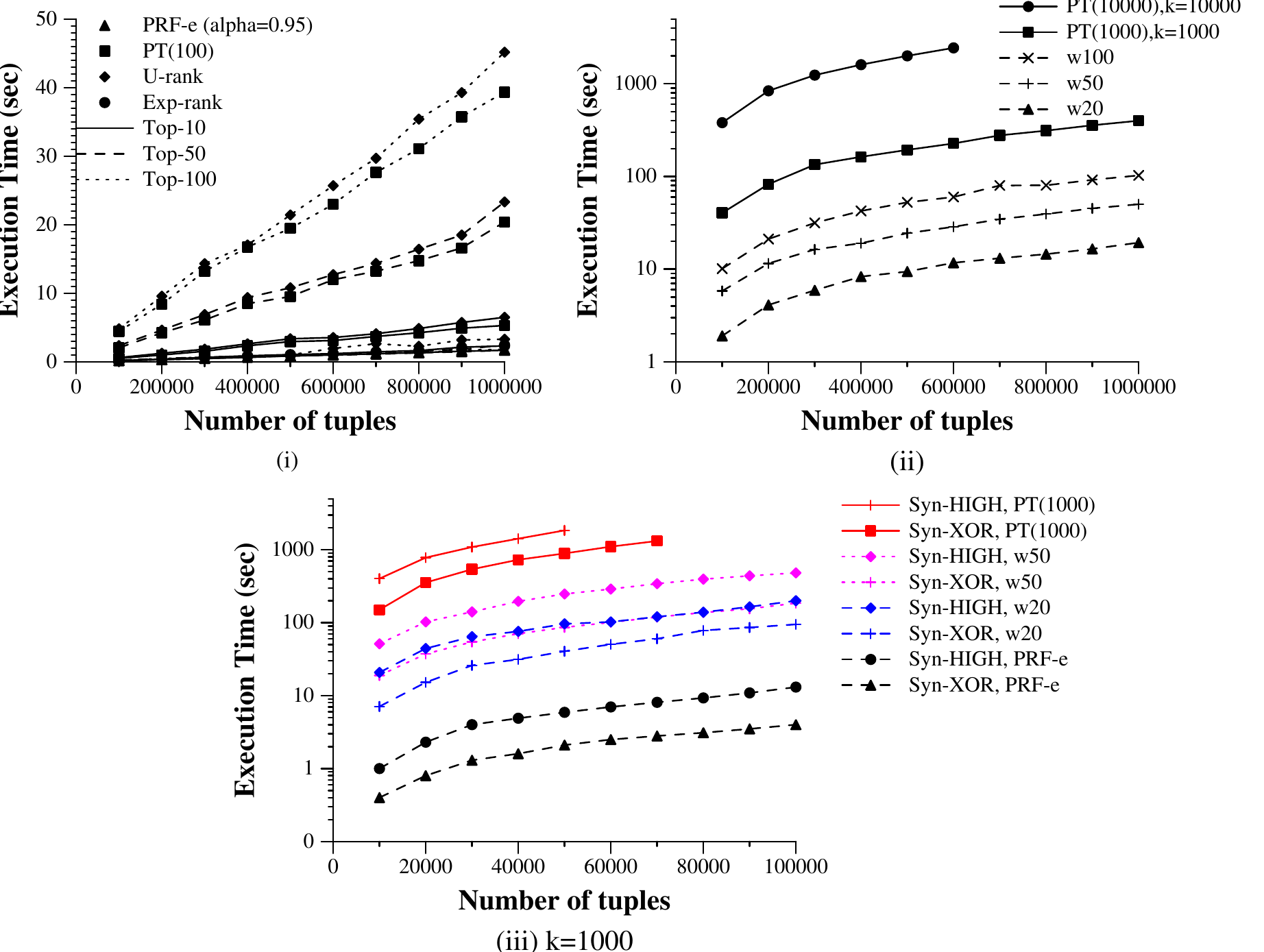}}
\vspace{-8pt}
\caption{Experiments comparing the execution times of the ranking algorithms (note that the y-axis is log-scale for (ii) and (iii))}
\label{fig:execution-times}
\end{figure*}

\subsection{Execution Times}
Figure \ref{fig:execution-times}(i) shows the execution times for four ranking functions: \PRFe, \PT, \URK\ and  \ERK,
for the IIP-datasets, for different dataset sizes and $\rmk$. We note that the running time for \PRFs\ is similar to that
of \PT. As expected, ranking by \PRFe\ or \ERK\ is very efficient (1000000 tuples can be ranked within 1 or 2 seconds).
Indeed, after sorting the dataset in an non-decreasing score order,
\PRFe\ needs only a single scan of the dataset, and \ERK\
needs to scan the dataset twice.
Execution times for $PT(h)$ and \URK-$\rmk$ increase linearly with $h$ and $\rmk$ respectively and the algorithms become
very slow for high $h$ and $\rmk$. The running times of both \PRFe\ and \ERK\ are not significantly affected by $\rmk$.

\eat{
These all indicate that the ranking obtained by these functions is essentially instructed by
the probabilities $\Prob(r(t)=i)$ which we use as features for learning and good approximations can be obtained by
using linear combination of these features.
However \ERK\ cannot be learnt at all in this range of sample size as seen in the figure.
This means \ERK\ ranking highly depends on some feature, the size of the dataset in this case,
which can not be characterized by probabilities.
}

Figure \ref{fig:execution-times}(ii) compares the execution time for \PT\ and
its approximation using a linear combination
of \PRFe\ functions (see Figure \ref{fig:approximating_2}(i)), for two different values of $k$.
$w50$ indicates that 50 exponentials were used in the
approximation (note that the approximate ranking, based on \PRFe, is insensitive to the value of $\rmk$). As we can see,
for large datasets and for higher values of $\rmk$,
exact computation takes several orders of magnitude more time to compute
than the approximation. For example, the exact algorithm takes nearly 1 hour for $n=500,000$ and $h=10,000$
while the approximate answer obtained using $L=50$ \PRFe\ functions takes only $24$ seconds and achieves a Kendall
distance $0.09$.

For correlated datasets, the effect is even more pronounced. In Figure \ref{fig:execution-times}(iii), we plot the results
of a similar experiment, but using two correlated datasets: Syn-XOR and Syn-HIGH. Note that the number of tuples in these
datasets is smaller by a factor of 10. As we can see, our generating functions-based algorithms for computing \PRFe\ are
highly efficient, even for datasets with high degrees of correlation.
As above, approximation of the \PT\ ranking function using
a linear combination of \PRFe\ functions is significantly cheaper to compute than using the exact algorithm.

Combined with the previous results illustrating that a linear combination of \PRFe\ functions can
approximate other ranking functions very well,
this validates the unified ranking approach that we propose in this paper.

\eat{
Figure \ref{fig:approx_experiment}(iii) shows the execution time of the approximation
(with running time $O(n\log(n)+nL)$
and the exact algorithm (with running time $O(n\rmk)$) for various $n$ and $h$.
Our approximated algorithm
is much superior to the exact algorithm in terms of running time
especially for very large $n$ and $h$.
}

\section{\PRF\ Computation for Arbitrary Correlations}
\label{sec:correlations}

Among many models for capturing the correlations in a probabilistic database,
graphical models (Markov or Bayesian networks) perhaps represent the most systematic approach~\cite{sen:vldbj09}.
The appeal of graphical models stems both from the pictorial representation of the
dependencies, and a rich literature on doing inference over them.
In this section, we present an algorithm for computing the \PRF\ function values for all tuples
of a correlated dataset when the correlations are represented using a graphical model. The resulting algorithm is
a non-trivial dynamic program over the {\em junction tree} of the graphical model.
Our main result is that we can compute the PRF function in polynomial time if the
junction tree of the graphical model has bounded treewidth.
It is worth noting that this result can not
subsume our algorithm for and/xor trees (Section \ref{subsec_andortree}) since the treewidth of the moralized graph of a
probabilistic and/xor tree may not be bounded.
\eat{Specifically our algorithm
can compute the PRF function in time  $O(2^{tw}n^2(2^{tw}+n)|\calT|)$, where $tw$ is
the treewidth of the graphical model (size of the largest clique in the junction tree)
and $|\calT|$ (which is always $\leq n$) is the number of cliques in the
junction tree.}
In some sense, this is close to {\em instance-optimal} since the complexity of the
underlying inference problem is itself
exponential in the treewidth of the graphical model (this however does not preclude the possibility that the ranking itself
could be done more efficiently without computing the PRF function explicitly -- however,
such an algorithm is unlikely to exist).

\subsection{Definitions}
We start with briefly reviewing some notations and definitions related to graphical models and junction trees.
Let $T=\{t_1,t_2,\ldots,t_n\}$ be the set of tuples in $D_T$, sorted in an non-increasing order of
their score values. 
For each tuple $t$ in $T$, we associate an indicator
random variable $X_t$, which is $1$ if $t$ is present, and $0$ otherwise.
Let $\calX=\{X_{t_1},\ldots,X_{t_n}\}$ and $\calX_i=\{X_{t_1},\ldots,X_{t_i}\}$.
For a set of variables $S$, we use $\Prob(S)$ to denote the joint probability distribution over those
variables. So $\Prob(\calX)$ denotes the joint probability distribution that we are trying to
reason about. This joint distribution captures all the correlations in the dataset. However, directly
trying to represent it would take $O(2^n)$ space, and hence is clearly infeasible.

Probabilistic graphical models allow us to represent this joint distribution compactly by
exploiting the conditional independences present among the variables. Given three disjoint sets of
random variables $A, B, C$, we say that $A$ is conditionally independent of $B$ {\em given} $C$ if and only if:
\[ \Prob(A, B | C) = \Prob(A | C) \Prob(B | C) \]

We assume that we are provided with a {\em junction tree} over the variables $\calX$ that captures
the correlations among them. A junction tree can be constructed from a graphical model using standard
algorithms~\cite{Jensen94}. Recently junction trees have also been used as a internal representation
for probabilistic databases, and have been shown to be quite effective at handling lightly correlated
probabilistic databases~\cite{conf/sigmod/KanagalD09}.
We describe the key properties of junction trees next.

\topic{Junction tree}
Let $\calT$ be a tree with each node $v$ associated with a subset $C_v\subseteq \calX$. We say $\calT$ is a {\em junction tree} if any intersection
$C_u\cap C_v$ for any $u,v\in \calT$ is contained in $C_w$ for every node $w$ on the unique path
between $u$ and $v$ in $\calT$ (this is called the {\em running intersection property}).
The treewidth $tw$ of a junction tree is defined to be $\max_{v\in \calT} |C_v|-1$.

Denote $S_{u,v}=C_v\cap C_u$
for each edge $(u,v)\in \calT$. We call $S_{u,v}$ a {\em separator} since removal of $S_{u,v}$
disconnects the graphical model. The set of conditional independences embodied by a junction
tree can be found using the Markov property: \\[2pt]
{\bf (Markov Property)} Given variable sets $A, B, C$, if $C$ separates $A$ and $B$ (i.e.,
removal of variables in $C$ disconnects the variables in $A$ from variables in $B$ in the
junction tree), then $A$ is conditionally independent of $B$ given $C$.

\begin{figure}[t]
\centerline{\includegraphics[width=0.6\textwidth]{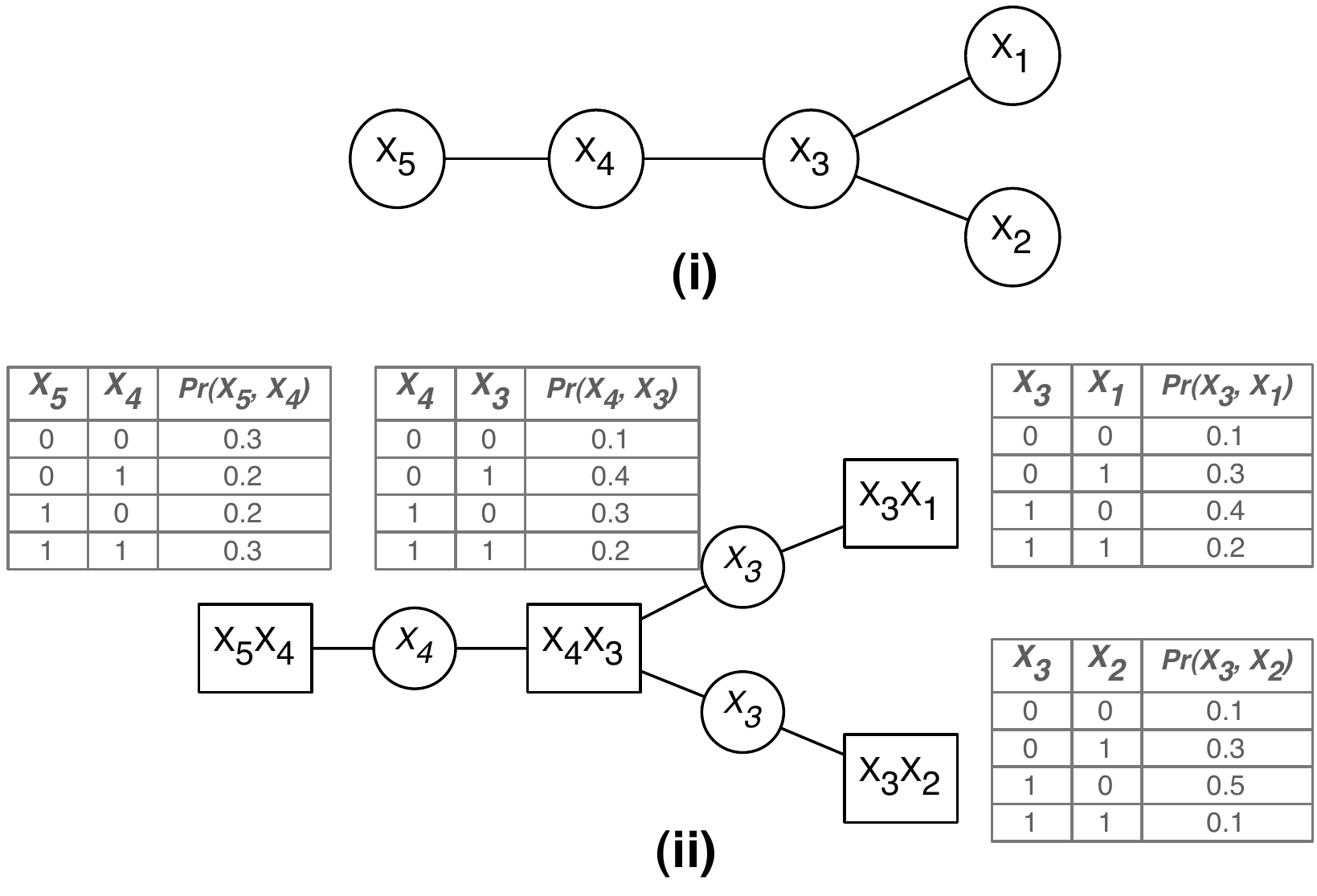}}
\caption{(i) A graphical model; (ii) A junction tree for the model along with the (calibrated) potentials.}
\label{fig:graphical-model}
\end{figure}

\begin{example}
Let $T=\{t_1,t_2,t_3,t_4,t_5\}$. Figure \ref{fig:graphical-model} (i) and (ii)
show the (undirected) graphical model 
and the corresponding junction tree $\calT$. 
$\calT$ has four nodes: 
$C_{1}=\{X_{t_4},X_{t_5}\}$, $C_{2}=\{X_{t_4},X_{t_3}\}$,
$C_{3}=\{X_{t_3},X_{t_1}\}$  and $C_{4}=\{X_{t_3},X_{t_2}\}$.
The treewidth of $\calT$ is $1$.
We have, $S_{1,2}=\{X_4\}$, $S_{2,3}=\{X_3\}$ and $S_{2,4}=\{X_3\}$.
Using the Markov property, we observe that $X_5$ is independent of $X_1, X_2, X_3$ given $X_4$.
\end{example}

\topic{Clique and Separator Potentials} With each clique $C_v$ in the junction tree, we associate a
{\em potential} $\pi_v(C_v)$,
which is a function over all variables $X_{t_i} \in C_v$  and
captures the correlations among those variables. 
Similarly, with each separator $S_{u,v}$, we associate
a {\em potential} $\mu_{u,v}(S_{u,v})$.
Without loss of generality, we assume that the
potentials are {\em calibrated}, i.e., the potential corresponding to a clique (or a separator) is
exactly the joint probability distribution over the variables in that clique (separator). Given
a junction tree with arbitrary potentials, calibrated potentials can be computed using a standard {\em message
passing algorithm}~\cite{Jensen94}. The complexity of this algorithm is $O(n 2^{tw})$.
Then the joint probability distribution of $\calX$,
whose correlations can be captured using a calibrated junction tree $\calT$, can be written as:
$$
\Prob(\calX)={\prod_{v\in \calT}\pi_v(C_v) \over \prod_{(u,v)\in \calT}\mu_{u,v}(S_{u,v})}
={\prod_{v\in \calT}\Prob(C_v) \over \prod_{(u,v)\in \calT}\Prob(S_{u,v})}
$$


\subsection{Problem Simplification}
We begin with describing the first step of our algorithm, and defining a reduced and
simpler to state problem.

Recall that our goal is to rank the tuples according to $\rank(t_i) = \sum_{j>0}\omega(j) \Prob(r(t_i)=j)$.
For this purpose, we first compute the positional probabilities, $\Prob(r(t_i) = j)$\ \  $\forall j\ \  \forall t_i$, using the algorithms
presented in the next two subsections. Given those, the values of $\rank(t_i)$ can be computed
in $O(n^2)$ time for all tuples, and the ranking itself can be done in $O(n\log(n))$ time (by sorting).
The positional probabilities ($Pr(r(t_i) = j)$) may also be of interest by themselves.

For each tuple $t_i$, we compute $\Prob(r(t_i) = j)\ \forall j$ at once. Recall that $\Prob(r(t_i) = j)$
is the probability that $t_i$ exists (i.e., $X_i = 1$) and exactly $j-1$ tuples with
scores higher than $t_i$ are present (i.e., $\sum_{l=1}^{i-1} X_l = j-1$). In other words:
\begin{eqnarray*}
\Prob(r(t_i) = j) &=& \Prob(X_i = 1 \wedge \sum_{l=1}^{i-1} X_l = j-1)  \\
 &=& \Prob( (\sum_{l=1}^{i-1} X_l = j-1) | X_i = 1) \Prob(X_i = 1)
\end{eqnarray*}

Hence, we begin with first conditioning the junction tree by setting $X_i = 1$, and re-calibrating.
This is done by identifying all cliques and separators which contain $X_i$, and by updating the
corresponding probability distributions by removing the values corresponding to $X_i = 0$.
More precisely, we replace a probability distribution $\Prob(X_{i_1}, \dots, X_{i_k}, X_i)$,
by a potential $\pi(X_{i_1}, \dots, X_{i_k})$ computed as:
\begin{eqnarray*}
&&\pi(X_{i_1} = v_1, \dots, X_{i_k} = v_k) \\
&&= \Prob(X_{i_1} = v_1, \dots, X_{i_k} = v_k, X_i = 1)
\end{eqnarray*}

\begin{figure}[h]
\centerline{\includegraphics[width=0.6\textwidth]{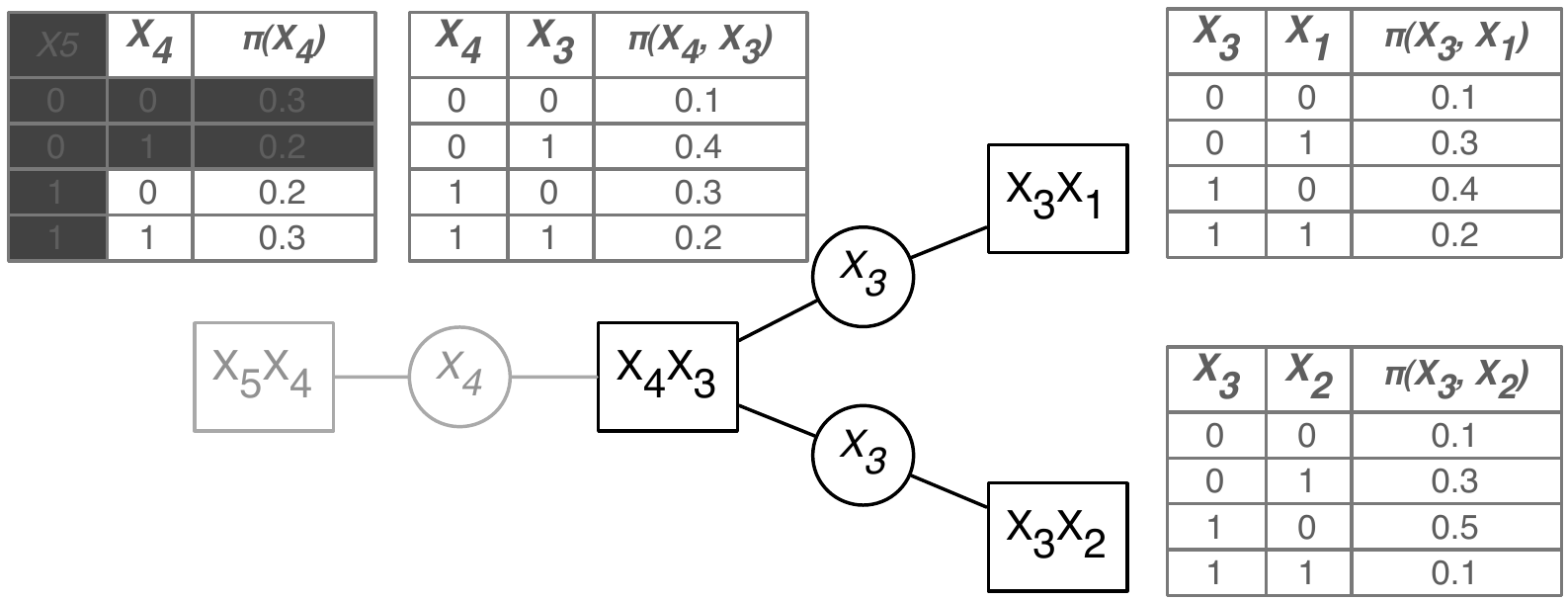}}
\caption{Conditioning on $X_5 = 1$ results in a smaller junction tree, with uncalibrated potentials, that captures the distribution over
$X_1, X_2, X_3, X_4$ given $X_5 = 1$. }
\label{fig:junction-tree-remove-5}
\vspace{-5pt}
\end{figure}

$\pi$ is not a probability distribution since the entries in it may not sum up to 1.
Further, the potentials may not be consistent with each other. Hence, we need to recalibrate
this junction tree using message passing~\cite{Jensen94}. As mentioned earlier, this takes $O(n 2^{tw})$ time.
Figure \ref{fig:junction-tree-remove-5} shows the resulting (uncalibrated) junction tree after conditioning
on $X_5 = 1$.

\begin{figure}[t]
    \vspace{-10pt}
\centerline{\includegraphics[width=0.16\textwidth]{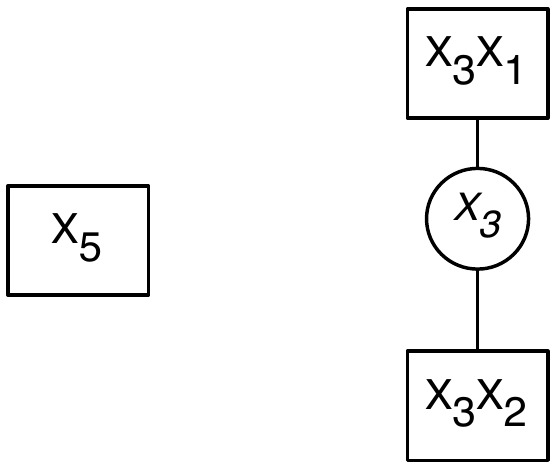}}
\caption{Conditioning on $X_4 = 1$ results in two junction trees.}
\label{fig:junction-tree-remove-4}
    \vspace{-5pt}
\end{figure}

If $X_i$ is a separator in the junction tree, then we get more than one junction tree
after conditioning on $X_i = 1$. Figure \ref{fig:junction-tree-remove-4}
shows the two junction trees we would get after conditioning on $X_4 = 1$.
The variables in these junction trees are independent of each other (this follows from
the Markov property), and the junction trees can be processed separately from each other.

Since the resulting junction tree or junction trees capture the probability distribution
conditioned on the event $X_i = 1$, our problem now reduces to finding the probability
distribution of $\sum_{l = 1}^{i-1} X_l$ in those junction trees.
For cleaner description of the algorithm, we associate an indicator variable $\delta_{X_l}$
with each variable $X_l$ in the junction tree. $\delta_{X_l}$ is set to 1 if $l \le i-1$,
and is 0 otherwise. This allows us to state the key problem to be solved as follows:

\smallskip
\smallskip
{\em
\noindent{\underline{\bfseries Redefined Problem}\footnote{We rename the variables to avoid confusion.}:} Given a junction tree over $m$ binary variables $Y_1, \dots, Y_m$,
where each variable $Y_j$ is associated with an indicator variable $\delta_{Y_j} \in \{0, 1\}$,
find the probability distribution of the random variable $\PS = \sum_{l = 1}^{m} Y_l \delta_l$.
}
\smallskip
\smallskip

If the result of the conditioning was a single junction tree (over $m = n-1$ variables), we multiply the
resulting probabilities by $\Prob(X_i = 1)$ to get the rank distribution of $t_i$.

However, if we get $k > 1$ junction trees, then we need one additional step.
Let $\PS_1, \dots, \PS_k$ be the random variables denoting the partial sums for each of
junction trees. We need to combine the probability distributions over these partial
sums, $\Prob(\PS_1), \dots, \Prob(\PS_k)$, into a single probability distribution over
$\Prob(\PS_1 + \dots + \PS_k)$. This can be done by repeatedly applying the following
general formula:
\[ \Prob(\PS_1 + \PS_2 = a) = \sum_{j = 0}^{a} \Prob(\PS_1 = j) \Prob(\PS_2 = a - j) \]

A naive implementation of the above takes time $O(n^2)$. Although this can be improved
using the ideas presented in Appendix \ref{app_expandpoly}, the complexity of computing
$\Prob(\PS_i)$ is much higher and dominates the overall complexity.

Next we present algorithms for solving the redefined problem.

\subsection{Algorithm for Markov Sequences}
\begin{figure}
\centerline{\includegraphics[width=0.6\linewidth]{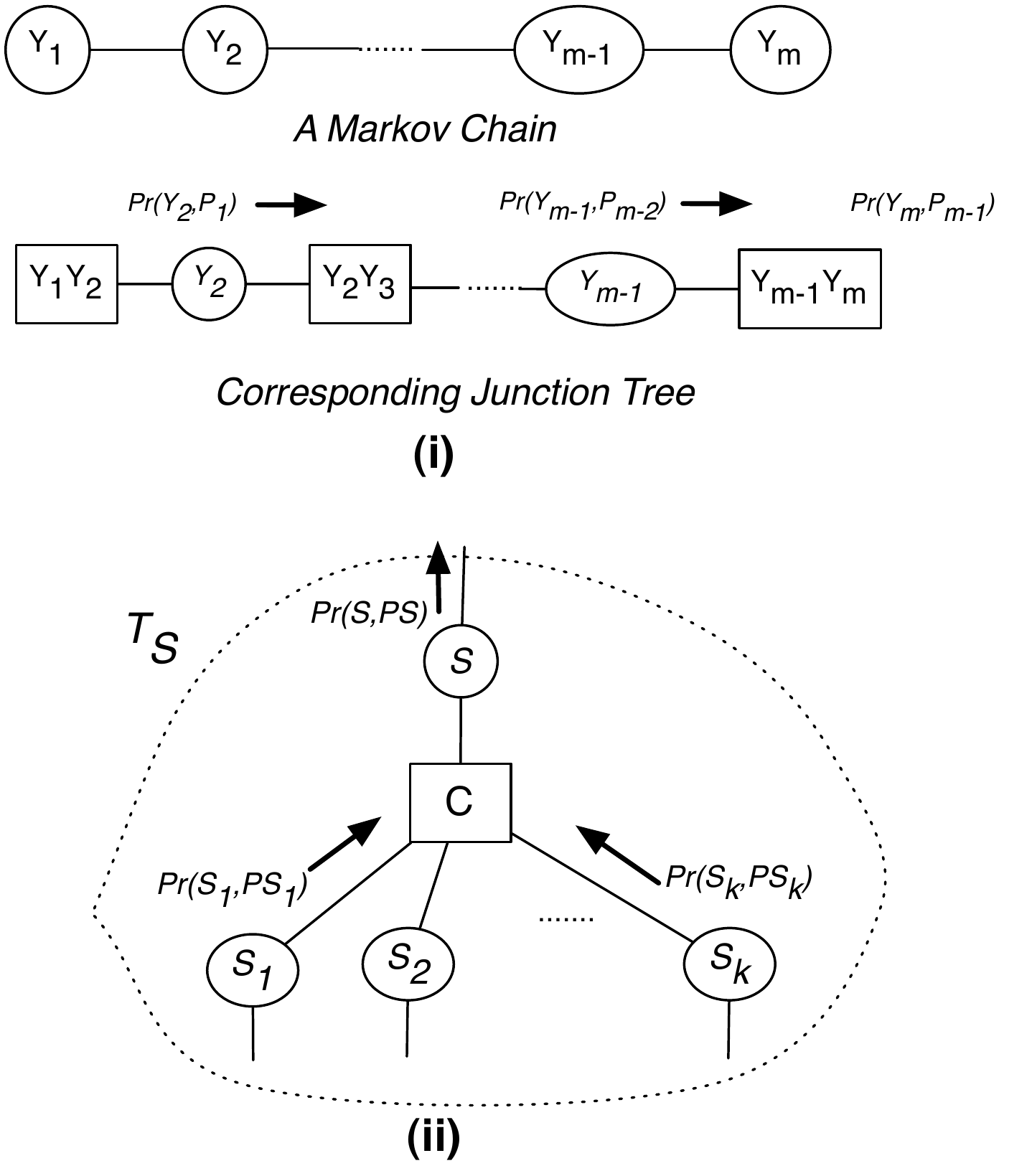}}
\vspace{-10pt}
\caption{(i) A Markov chain, and the corresponding junction tree; (ii) Illustrating the recursion
for general junction trees.}
\vspace{-5pt}
\label{fig:general-junction-tree}
\end{figure}

We first describe an algorithm for Markov chains, a special, yet important, case of the graphical models.
Markov chains appear naturally in many settings, and have been studied in probabilistic database literature
as well~\cite{kanagal:icde09,DBLP:conf/sigmod/ReLBS08,conf/pods/KimelfeldR10}.
Any finite-length Markov chain is a Markov network whose underlying graph
is simply a path: each variable is directly dependent on only its predecessor
and successor. The junction tree for a Markov chain is also a path in which each node
corresponds to an edge of the Markov chain. The treewidth of such a junction tree
is one.  Without loss of generality, we
assume that the Markov chain is $Y_1, \dots, Y_m$ (Figure \ref{fig:general-junction-tree}(i)).
The corresponding junction tree $\calT$ is a path with cliques $C_j = \{Y_{j},Y_{j+1}\}$
as shown in the figure.

We compute the distribution $\Prob(\sum_{l = 1}^{m} Y_l \delta_l)$ recursively.
Let $\PS_j = \sum_{l = 1}^{j} Y_l \delta_l$ denote the
partial sum over the first $j$ variables $Y_1, \dots, Y_j$.

At the clique $\{Y_{j-1}, Y_{j}\}$, $j \ge 1$, we recursively compute the joint probability distribution: $\Prob(Y_{j}, \PS_{j-1})$.
The initial distribution $\Prob(Y_2, \PS_1)$, $\PS_1 = \delta_1 Y_1$, is computed directly:
\begin{eqnarray*}
\Prob(Y_2, \PS_1 = 0) &=& \Prob(Y_2, Y_1 = 0) + (1 - \delta_i) \Prob(Y_2, Y_1 = 1) \\
\Prob(Y_2, \PS_1 = 1) &=& \delta_i \Prob(Y_2, Y_1 = 1).
\end{eqnarray*}
Given $\Prob(Y_j, \PS_{j-1})$, we compute $\Prob(Y_{j+1}, \PS_j)$ as follows.
Observe that $\PS_{j-1}$ and $Y_{j+1}$ are conditionally independent given the value of
$Y_j$ (by Markov property). Thus we have:
\[ \Prob(Y_{j+1}, Y_{j}, \PS_{j-1}) = {\Prob(Y_{j+1}, Y_{j}) \Prob(Y_j, \PS_{j-1}) \over \Prob(Y_j)} \]
Using $\Prob(Y_{j+1}, Y_{j}, \PS_{j-1})$, we can compute:
\begin{eqnarray*}
 \Prob(Y_{j+1}, \PS_j = a) &=& \Prob(Y_{j+1}, Y_j = 0, \PS_{j-1} = a) \\
 && + \Prob(Y_{j+1}, Y_j = 1, \PS_{j-1} = a - \delta_j)
\end{eqnarray*}

\noindent{At} the end, we have the joint distribution: $\Prob(Y_{m}, \PS_{m-1})$. We can compute a distribution
over $\PS_m$ as:
\begin{eqnarray*}
 \Prob(\PS_m = a) &=& \Prob(Y_m = 0, \PS_{m-1} = a) \\
 && + \Prob(Y_m = 1, \PS_{m-1} = a - \delta_m)
\end{eqnarray*}

\topic{Complexity} The complexity of the above algorithm to compute $\Prob(\PS_m)$ is $O(m^2)$ --
although we only perform $m$ steps, $\Prob(Y_{j+1}, \PS_j)$ contains $2(j+1)$ terms, 
each of which
takes $O(1)$ time to compute.
Since we have to repeat this for every tuple, the overall complexity of ranking the dataset can be seen
to be $O(n^3)$.

\subsection{General Junction Trees}
We follow the same general idea for general junction trees. 
Let $\calT$ denote the junction tree over the variables $\calY = \{Y_1, \dots, Y_m\}$. We begin by rooting $\calT$ at
an arbitrary clique, and recurse down the tree. For a separator $S$, let $\calT_S$ denote the subtree
rooted at $S$. Denote by $\PS_S$ the partial sum over the variables in the subtree $\calT_S$ that are {\em not present}
in $S$, i.e.,:
\[ \PS_S = \sum_{j \in \calT_S, j \notin S} \delta_j X_j \]


Consider a clique node $C$, and let $S$ denote the separator between $C$ and its parent node ($S = \phi$ for the root clique node).
We will recursively compute the joint probability distribution $\Prob(S, \PS_S)$ for each such separator $S$. Since the root clique node has
no parent, at the end we are left with precisely the probability distribution that we need, i.e., $\Prob(\sum_{j=1}^m Y_i \delta_i$).

\topic{$C$ is an interior or root node} Let the separators to the children of $C$ be $S_1, \dots, S_k$ (see Figure \ref{fig:general-junction-tree}(ii)).
We recursively compute $\Prob(S_i, \PS_{S_i}), i = 1, \dots, k$.

Let $Z = C \setminus S$. We observe that $Z$ is precisely the set of variables
that contribute to the partial sum $\PS_S$, but do not contribute to any of the partial sums $\PS_{S_1}, \dots, \PS_{S_k}$, i.e.:
\[ \PS_{S} = \PS_{S_1} + \dots + \PS_{S_k} + \sum_{Z_i \in Z} \delta_{Z_i} Z_i \]

\noindent{We} begin with computing $\Prob(C, \PS_{S_1} + \dots + \PS_{S_k})$.
Observe that the variable set $C \setminus S_1$ is independent of $\PS_{S_1}$ given the values of the variables in $S_1$
(by Markov property). Note that it was critical that the variables in $S_1$ not contribute to the partial sum $\PS_{S_1}$,
otherwise this independence would not hold. Given that, we have:
\begin{eqnarray*}
\Prob(C, \PS_{S_1}) &=& \Prob(C \setminus S_1, S_1, \PS_{S_1})  \\
&=& {\Prob(C \setminus S_1, S_1) \Prob(S_1, \PS_{S_1}) \over \Prob(S_1)}
\end{eqnarray*}

\noindent{Using} $\PS_{S_2}$ is independent of $C \cup \{\PS_{S_1}\}$ given $S_2$, we get:
\[
\Prob(C, \PS_{S_1}, \PS_{S_2}) = {\Prob(C, \PS_{S_1}) \Prob(S_2, \PS_{S_2}) \over \Prob(S_2)}
\]
Now we can compute the probability distribution over $\Prob(C, \PS_{S_1} + \PS_{S_2})$ as follows:
{\small
\begin{eqnarray*}
&&\Prob(C, \PS_{S_1} + \PS_{S_2} = a) = \sum_{j = 0}^{a} \Prob(C, \PS_{S_1} = j, \PS_{S_2} = a - j)\\
&&= \sum_{j = 0}^{a} {\Prob(C, \PS_{S_1} = j) \Prob(S_2, \PS_{S_2} = a -j) \over \Prob(S_2)}
\end{eqnarray*}
}
By repeating this process for $S_3$ to $S_k$, we get the probability distribution: $\Prob(C, \PS_{S_1} + \dots + \PS_{S_k})$.

Next, we need to add in the contributions of
the variables in $Z$ to the partial sum $\PS_{S_1} + \dots + \PS_{S_k}$.
Let $Z$ contain $l$ variables, $Z_1, \dots Z_l$, and let $\delta_{Z_1}, \dots, \delta_{Z_l}$
denote the corresponding indicator variables.
It is easy to see that: \\[1pt]
{\small
\begin{eqnarray*}
    && \Prob(C \setminus Z, Z_1 = v_1, \dots, Z_k = v_k, \sum_{j=1}^k \PS_{S_j} + \sum_{j = 1}^l \delta_{z_j} Z_j = a) \\
    && = \Prob(C \setminus Z, Z_1 = v_1, \dots, Z_k = v_k, \sum_{j=1}^k \PS_{S_j} = a - \sum_{l = 1}^l \delta_{z_j} Z_j)
\end{eqnarray*}
}
where $v_i \in \{0, 1\}$.
Although it looks complex, we only need to touch every entry of the probability distribution $\Prob(C, \PS_1 + \dots + \PS_k)$ once to compute $\Prob(C, \PS_S)$.

All that remains is marginalizing that distribution to sum out the variables in $C \setminus S$, giving us
$\Prob(S, \PS_S)$.

\topic{$C$ is a leaf node (i.e., $k = 0$)} This is similar to the final step above. Let $Z = C \setminus S$ denote the
variables that contribute to the partial sum $\PS_S$. We can apply the same procedure as above to compute $\Prob(C, \PS_S = \sum_{Z_i \in Z} \delta_{Z_i} Z_i)$,
which we marginalize to obtain $\Prob(S, \PS_S)$.

\topic{Overall Complexity}
The complexity of the above algorithm for a specific clique $C$ is dominated by the cost of computing
the different probability distributions of the form $\Prob(C, \PS)$, where $\PS$ is a partial sum.
We have to compute $O(n)$ such probability distributions, and each of those computations takes $O(n^2 2^{|C|})$
time. Since there are at most $n$ cliques, and since we have to repeat this process for every tuple, the overall complexity
of ranking the dataset can be seen to be: $O(n^4 2^{tw})$,
where $tw$ denotes the treewidth of the junction tree, i.e., the size of the maximum clique minus 1.

\vspace{-0.2cm}
\section{Conclusions}
In this article we presented a unified framework for ranking over
probabilistic databases, and presented several novel and highly efficient algorithms for answering
\Topk\ queries. Considering the complex interplay between probabilities and scores, instead of
proposing a specific ranking function, we propose using two parameterized ranking functions,
called \PRFs\ and \PRFe, which allow the user to control the tuples that appear in the \Topk\ answers.
We developed novel algorithms for evaluating these ranking functions over large, possibly correlated,
probabilistic datasets. We also developed an approach for approximating a ranking function using
a linear combination of \PRFe\ functions thus enabling highly efficient, albeit approximate computation, and also for learning a ranking function from user preferences.

Our work opens up many avenues for further research. 
There may be other non-trivial subclasses of \PRF\ functions, aside
from \PRFe, that can be computed efficiently.
Understanding the behavior of various ranking functions and their relationships
across probabilistic databases with diverse uncertainties and
correlation structures also remains an important open problem in this area. Finally, the issues of ranking
have been studied for many years in disciplines ranging from economics to information retrieval; better understanding
the connections between that work and ranking in probabilistic databases remains a fruitful direction for further research.

\vspace{-0.2cm}
{

}

\appendix


\vspace{-0.3cm}
\section{Proofs}
\label{sec:proofs}

\noindent{\bf Theorem \ref{thm_generating}} {\em
The coefficient of the term $\prod_{j}x_j^{i_j}$ in $\calF(\calX)$ is the total probability of
the possible worlds for which, for all $j$, there are exactly $i_j$ leaves associated with variable $x_j$.
}

\begin{proof}
Suppose $\calT$ is rooted at $r$, $r_1,\ldots,r_h$ are $r$'s children, and $\calT_l$
is the subtree rooted at $r_l$.
We denote by $S$ (or $S_l$) the random set of leaves generated according to model $\calT$ (or $\calT_l$).
We let $\calF$ (or $\calF_{l}$) be the generating function corresponding to $\calT$ (or $\calT_{l}$).
For ease of notation, we use $\bfi$ to denote index vector $\langle i_{1},i_{2},\ldots\rangle$, $I$ to denote
the set of all such $\bfi$s
and $\calX^{\bfi}$ to denote $\prod_{j}x_{j}^{i_{j}}$.
Therefore, we can write
$\calF(\calX)=\sum_{i_1,i_2,\ldots}c_{i_1,i_2\ldots}x_1^{i_1}x_2^{i_2}\ldots=\sum_{\bfi\in I}c_{\bfi}\calX^{\bfi}.$
We use the notation $S\cong \bfi$ for some $\bfi=\langle i_{1},i_{2},\ldots \rangle\in I$ to denote
the event that $S$ contains $i_{j}$ leaves associated with variable $x_{j}$ for all $j$.
Given the notations, we need to show $c_{\bfi}=\Prob(S\cong \bfi)$.

We shall prove by induction on the height of the and/xor tree.
We consider two cases.
If $r$ is a $\Cwedge$ node, we know from Definition~\ref{and/xor} that
$S=\cup_{l=1}^h S_l$.
First, it is not hard to see that given $S_{l}\cong \bfi_{l}$ for $1\leq l\leq h$,
the event $S\cong \bfi$ happens if and only if $\sum_{l}\bfi_{l}=\bfi$.
Therefore,
\begin{align}
\label{eq:prob-subset}
\Prob(S\cong \bfi)
=\sum_{\sum_{l} \bfi_{l}=\bfi} \prod_{l=1}^{h} \Prob(S_l \cong \bfi_{l}).
\end{align}
Assume $\calF_{l}$ can be written as $\sum_{\bfi_{l}} c_{l,\bfi_{l}}\calX^{\bfi_{l}}$.
From the construction of the generating function,
we know that
\begin{align}
\label{eq:ind-gene}
\calF(\calX)&=\prod_{l=1}^{h}\calF_{l}=\prod_{l=1}^{h}\sum_{\bfi_{l}\in I} c_{l,\bfi_{l}}\calX^{\bfi_{l}}
=\sum_{\bfi\in I} \Bigl(\sum_{\sum_{l}\bfi_{l}=\bfi} \prod_{l=1}^{h} c_{l,\bfi_{l}} \calX^{\bfi_{l}} \Bigr)\notag \\
&=\sum_{\bfi\in I} \Bigl(\sum_{\sum_{l}\bfi_{l}=\bfi} \prod_{l=1}^{h} c_{l,\bfi_{l}} \Bigr) \calX^{\bfi}
\end{align}
By induction hypothesis, we have $\Prob(S_{l}\cong \bfi_{l})=c_{l,\bfi_{l}}$ for any $l$ and $\bfi_{l}$.
Therefore, we can conclude from (\ref{eq:prob-subset}) and (\ref{eq:ind-gene})
that $\calF(\calX)=\sum_{\bfi} \Prob(S\cong \bfi) \calX^{\bfi}$.

Now let us consider the other case where $r$ is a $\Cvee$ node.
From Definition~\ref{and/xor}, it is not hard to see that
\begin{align}
\label{eq:prob-subset2}
\Prob(S\cong \bfi)=\sum_{l=1}^{h}\Prob(S_l=\bfi)p_{(r,r_l)}
\end{align}
Moreover, we have
\begin{align*}
\calF(\calX)&=\sum_{l=1}^{h}p_{(r,r_l)}\calF_{l}(\calX)=\sum_{l=1}^{h}p_{(r,r_{l})}\sum_{\bfi_{l}} c_{l,\bfi_{l}} \calX^{\bfi_{l}} \notag\\
&=\sum_{\bfi}\Bigl(\sum_{l=1}^{h}p_{(r,r_{l})} c_{l,\bfi}\Bigr) \calX^{\bfi}=\sum_{\bfi} \Prob(S\cong \bfi) \calX^{\bfi}
\end{align*}
where the last equality follows from (\ref{eq:prob-subset2}) and induction hypothesis.
This completes the proof. \qed
\end{proof}

\section{Expanding Polynomials}
\label{app_expandpoly}
This section is devoted to several algorithms for expanding polynomials into standard forms.

\subsection{Multiplication of a Set of Polynomials}
\label{app_runningtime}

Given a set of polynomials in the form of $P_i=\sum_{j\geq 0}c_{ij}x^j$ for $1\leq i\leq k$, we want to compute
the multiplication $P=\prod_{i=1}^k P_i$ written in the standard form $P=\sum_{j\geq 0}c_jx^j$, i.e.,
we need to compute the coefficients $c_j$.
Let $d(P_i)$ be the degree of the polynomial $P_i$.
Let $n=\sum_i^k d(P_i)$ be the degree of $P$.

\topic{Naive Method}
First we note that the naive method (multiply $P_i$s one by one) gives us an $O(n^2)$ time algorithm by
simple counting argument.
Let $\bar{P}_i=\prod_{j=1}^i P_j$. It is easy to see $d(\bar{P}_i)=\sum_{j=1}^i d(P_i)$.
So the time to multiply $\bar{P}_i$ and $P_{i+1}$ is $O(d(\bar{P}_i)\cdot d(P_{i+1}))$.
Then, we can see the total time complexity is:
\begin{equation*}
\sum_{i=1}^{k-1} O(d(\bar{P}_i)\cdot d(P_{i+1}))=O(n)\cdot \sum_{i=1}^{k-1} d(P_{i+1})=O(n^2).
\end{equation*}

\topic{Divide-and-Conquer}
Now, we show how to use divide-and-conquer and FFT (Fast Fourier Transformation) to achieve an
$O(n\log^2 n)$ time algorithm. It is well known that the multiplication of two polynomials
of degree $O(n)$ can be done in $O(n\log n)$ time using FFT.
The divide-and-conquer algorithm is as follows:
If there exists any $P_i$ such that $d(P_i)\geq {1\over 3}d(P)$, we evaluate
$\prod_{j:j\ne i}P_i$ recursively and then multiply it with $P_i$ using FFT.
If not, we partition all $P_i$s into two sets $S_1$ and $S_2$
such that ${1\over 3}d(P)\leq d(\prod_{i\in S_i}P_i)\leq {2\over 3}d(P)$.
Then we evaluate $S_1$ and $S_2$ separately and multiply them together using FFT.
It is easy to see the time complexity of the algorithm running on input size $n$ satisfies
$$
T(n)\leq \max\{T({2\over 3}n)+O(n\log n), T(n_1)+T(n_2)+O(n\log n)\}
$$
where $n_1+n_2=n$ and ${1\over 3}n\leq n_1\leq n_2\leq {2\over 3}n$.
By solving the above recursive formula, we know $T(n)=O(n\log^2 n)$.

\subsection{Expanding a Nested Formula}
\label{app_runningtime2}

We consider a more general problem of
expanding a nested expression of uni-variable polynomial (with variable $x$) into its standard form
$\sum c_{i}x^{i}$.
Here a nested expression refers to a formula that only involves constants, the variable $x$,
addition $+$, multiplication $\times$, and parenthesis $($ and $)$, for example,
$f(x)=((1+x+x^2)(x^2+2x^3)+x^3(2+3x^4))(1+2x)$.
Formally, we define recursively an {\em expression} to be either
\vspace{-0.1cm}
\begin{enumerate}
\item[1.] A constant or the variable $x$, or
\item[2.] The sum of two {\em expressions}, or
\item[3.] The product of two {\em expressions}.
\end{enumerate}
\vspace{-0.1cm}
We assume the degree of the polynomial and the length of the expression are of sizes $O(n)$.

The naive method runs in time $O(n^3)$ (each inner node needs $O(n^{2})$ time as shown in the last subsection).
If we use the previous divide-and-conquer method for expanding
each inner node, you can easily get $O(n^2\log^2n)$.
Now we sketch two improved algorithms with running time $O(n^{2})$.
The first is conceptual simpler while the second is much easier to implement.

\topic{Algorithms 1}
\begin{enumerate}
\item Choose $n+1$ different numbers $x_0, ...., x_{n}$ .
\item Evaluate the polynomial at these points, i.e., compute $f(x_i)$.
It is easy to see that each evaluation takes linear time (bottom-up over the tree). So this step takes $O(n^2)$ time in total.
\item Use any $O(n^2)$ polynomial interpolation algorithm to find the coefficient.
In fact, the interpolation reduces to finding a solution for the following linear system:
$$
\begin{bmatrix}
x_0^n & x_0^{n-1} & x_0^{n-2} & \ldots & x_0 & 1 \\
x_1^n & x_1^{n-1} & x_1^{n-2} & \ldots & x_1 & 1 \\
\vdots & \vdots & \vdots & & \vdots & \vdots \\
x_n^n & x_n^{n-1} & x_n^{n-2} & \ldots & x_n & 1
\end{bmatrix}
\begin{bmatrix}
c_n \\
c_{n-1} \\
\vdots \\
c_0
\end{bmatrix}
=
\begin{bmatrix}
f(x_0) \\
f(x_1) \\
\vdots \\
f(x_n)
\end{bmatrix}.
$$
The commonly used Gaussian elimination for inverting a matrix requires $O(n^{3})$ operations.
The matrix we used is a special type of matrix and is commonly referred to as a Vandermonde matrix.
There exists numerical algorithms that can invert a Vandermonde matrix in $O(n^{2})$ time, for example \cite{Bjorck70}.
\end{enumerate}

A small drawback of the above algorithm is that the algorithms used to invert a Vandermonde matrix
is nontrivial to implement. The next algorithm does not need to invert a matrix, is much simpler to implement
and has the same running time of $O(n^{2})$.

\topic{Algoirthm 2}
We need some notation first.
Suppose the polynomial is $f(x)=\sum_{j=0}^{n} c_j x^j$ ($c_{j}$s are unknown yet).
Let $\bfe_i$ be the $(n+1)$-dimensional zero vector except that the $i^{\text{th}}$ entry is $1$, i.e., $\bfe_i=\langle 0,0,..,1,...,0,0\rangle$.
Let $\bfd_i=\langle1, e^{\frac{2\pi \uimag}{n+1}i},e^{\frac{2\pi \uimag}{n+1}2i},\ldots\rangle$ be the $n+1$-dimensional vector
which is the DFT (Discrete Fourier Transformation) of $\bfe_i$.
Let $u=e^{-\frac{2\pi\uimag}{n+1}}$ be the $n+1^{\text{th}}$ root of unit.
Let $\bfu=\langle 1,u,u^2,....,u^{n}\rangle$
and $\bfu^k=\langle 1,u^k,u^{2k},u^{3k},.....\rangle$.

By definition, $\bfe_i=\frac{1}{n+1}\sum_k \bfd_{ik} \bfu^k$ where $\bfd_{ik}$ is the $k^{\text{th}}$ entry of $\bfd_{i}$.
Let $\bfc=\langle c_{0},\ldots, c_{n}\rangle$ be the coefficient vector of $f$.
It is trivial to see $c_i = \bfc \cdot \bfe_i$ (the inner product).
Therefore, we have that
\begin{align}
\label{eq:poly}
c_{i}&=\bfc\cdot\bfe_i =\frac{1}{n+1}\sum_k \bfd_{ik} (\bfc\cdot \bfu^k)=\frac{1}{n+1}\sum_k \bfd_{ik} f(u^{k}).
\end{align}
The last equality holds by the definition of $f(x)$.
If we use $\mathbf{f}$ to denote the vector $\langle f(u^{0}),\ldots, f(u^{n})\rangle$
and $\mathbf{D}$ to denote the matrix $\{\bfd_{ij}\}_{0\leq,i,j\leq n}$,
the above equation can be simply written as
$$
\bfc={1\over n+1}\mathbf{D}\mathbf{f}.
$$
Now, we are ready describe our algorithm:
\begin{enumerate}
\item Compute $f(u^k)$ for all $k$. This consists of evaluating $f(x)$ over complex $x$ $n$ times, which takes $O(n^2)$ time.
\item Use (\ref{eq:poly}) to compute the coefficients. This again takes $O(n^{2})$ time.
\end{enumerate}

In fact, the above algorithm can be seen as a specialization of the first algorithm.
Instead of picking arbitrary $n+1$ real points $x_{0},\ldots x_{n}$ to evaluate the polynomial,
we pick $n+1$ complex points $1, u, u^{2}, \ldots, u^{n}$.
The Vandermonde matrix formed by these points, i.e.,
$$\bfF =
\begin{bmatrix}
 u^{0 \cdot 0}     & u^{0 \cdot 1}     & \ldots & u^{0 \cdot n}     \\
 u^{1 \cdot 0}     & u^{1 \cdot 1}     & \ldots & u^{1 \cdot n}     \\
 \vdots                   & \vdots                   & \ddots & \vdots                       \\
 u^{n \cdot 0} & u^{n \cdot 1} & \ldots & u^{n \cdot n} \\
\end{bmatrix}
$$
has a very nice property that
$$
\bfF^{-1}=\frac{1}{n+1}\bfF^*
$$
where $\bfF^{*}$ is the conjugate of $\bfF$ (This can be verified easily).
Therefore, we can obtain $\bfF^{-1}$ for free.
Actually, it is easy to see that $\bfF^{*}$ is exactly $\mathbf{D}$.



\end{document}